\documentclass[11pt]{article}
\usepackage[utf8]{inputenc}
\pdfoutput = 1

\usepackage{amsmath}
\usepackage{amssymb}
\usepackage{bbm}
\usepackage{bbold}
\usepackage{braket}
\usepackage{caption}
\usepackage{cite}
\usepackage{color}
    \definecolor{darkgreen}{rgb}{0,0.5,0}
    \definecolor{darkred}{rgb}{0.5,0,0}
    \definecolor{darkblue}{rgb}{0,0,0.6}
    \definecolor{purple}{rgb}{0.4,.2,0.7}
\usepackage{float}
\usepackage{graphicx}
\usepackage[hyperfootnotes = true, colorlinks = true, linkcolor = darkblue, citecolor = purple]{hyperref}
\usepackage{mathabx}
\usepackage[mathcal]{eucal}
\usepackage{pdfsync}
\usepackage{url}
\usepackage[normalem]{ulem}
\usepackage{comment}
\usepackage[english]{babel}
\usepackage{mathtools,slashed}
\usepackage{csquotes}
\usepackage{subcaption}

\numberwithin{equation}{section}

\usepackage[margin = 2.2cm]{geometry}
\usepackage[ragged]{footmisc}
\renewcommand{\d}{\mathrm{d}}
\renewcommand{\i}{\mathrm{i}}

\newcommand{\TT}{\mathcal{T}}

\DeclareMathOperator{\Tr}{Tr}

\DeclareMathOperator{\asin}{arcsin}
\DeclareMathOperator{\acos}{arccos}
\DeclareMathOperator{\atan}{arctan}
\DeclareMathOperator{\sech}{sech}

\DeclareMathOperator{\atanh}{arctanh}
\DeclareMathOperator{\sgn}{sgn}

\renewcommand{\Re}[1]{{\mathrm{Re}}\left[ #1 \right]}

\newcommand{\rA}{A}
\newcommand{\LR}{\ell}

\begin{document}

\vspace{1.2cm}
\begin{center}
\noindent{\bf \LARGE Holographic BCFT with a Defect on the}

\noindent{\bf \LARGE End-of-the-World Brane}

\vspace{0.4cm}

{\bf \large Masamichi Miyaji$^{1,2}$ and Chitraang Murdia$^{3,4}$}
\vspace{0.3cm}\\

{\it $^1$ Institute for Advanced Research, Nagoya University,\\ Nagoya, Aichi 464-8601, Japan}\\ 
{\it $^2$Department of Physics, Nagoya University,\\ Nagoya, Aichi 464-8602, Japan}\\

{\it $^3$ Berkeley Center for Theoretical Physics, Department of Physics, \\
University of California, Berkeley, CA 94720, USA}\\
{\it $^4$ Theoretical Physics Group, Lawrence Berkeley National Laboratory, \\ Berkeley, CA 94720, USA}\\

\end{center}

\begin{abstract}

In this paper, we propose a new gravity dual for a $2$d BCFT with two conformal boundaries by introducing a defect that connects the two End-of-the-World branes. 
We demonstrate that the BCFT dual to this bulk model exhibits a richer lowest spectrum.
The corresponding lowest energy eigenvalue can continuously interpolate between $-\frac{\pi c}{24\Delta x}$ and $0$ where $\Delta x$ is the distance between the boundaries.
This range was inaccessible to the conventional AdS/BCFT model with distinct boundary conditions. 
We compute the holographic entanglement entropy and find that it exhibits three different phases, one of which breaks the time reflection symmetry.
We also construct a wormhole saddle, analogous to a $3$d replica wormhole, which connects different boundaries through the AdS bulk. 
This saddle is present only if the BCFT is non-unitary and is always subdominant compared to the disconnected saddle. 

\end{abstract}

\tableofcontents

\section{Introduction}

AdS/BCFT \cite{Karch:2000gx, Takayanagi:2011zk, Fujita:2011fp} studies the gravity dual of boundary conformal field theory \cite{Cardy:1984bb, Cardy:2004hm}.
The simplest, bottom-up model proposed for AdS/BCFT is a spacetime terminating at the End-of-the-World (EOW) brane \cite{Takayanagi:2011zk, Fujita:2011fp, Nozaki:2012qd}.
This EOW brane is anchored to the BCFT boundary.
This bottom-up model captures some qualitative aspects of stringy models for AdS/BCFT \cite{Chiodaroli:2011fn, Aharony:2011yc, Chiodaroli:2012vc}, and has served as a useful toy model for black hole evaporation  \cite{Penington:2019npb, Almheiri:2019psf, Almheiri:2019hni, Penington:2019kki, Almheiri:2019qdq, Rozali:2019day, Chen:2019uhq, Geng:2020qvw, Chen:2020uac, Chen:2020hmv, Kruthoff:2021vgv, Afrasiar:2022ebi}, interpreted as doubly-holographic brane-world models \cite{Randall:1999ee, Randall:1999vf, Karch:2000ct, Giddings:2000mu}.
On the other hand, this simplest model of AdS/BCFT is known to have several atypical features in the boundary operator spectrum amongst all holographic BCFTs, such as fine-tuned boundary operator spectrum \cite{Reeves:2021sab} and the absence of interactions between distinct EOW branes. The latter, which is the main focus of this paper, results in a fixed gap between the lowest eigenvalues of the BCFT with two distinct boundaries and the BCFT with identical boundaries \cite{Miyaji:2021ktr}. 

In this paper, we modify the conventional AdS/BCFT model by allowing the EOW branes with different tensions to be connected at defects.
Geometries with intersecting EOW branes were first considered in \cite{Geng:2021iyq}.
In our model, we treat the defect explicitly by including an additional contribution to the action from the intersection.
Clearly, this defect leads to an interaction between the EOW branes.
We find that this interaction leads to several novel results:

\begin{itemize}

    \item The defect modifies the lowest eigenvalue of the conventional AdS/BCFT model.
    In fact, the lowest eigenvalue in our refined model continuously interpolates between the identical boundary BCFT and the aforementioned spectral gap.
    We also include a bulk conical defect on the gravity side, which corresponds to a boundary operator \cite{Kawamoto:2022etl, Kusuki:2021gpt, Geng:2021iyq, Kusuki:2022wns, Kusuki:2022ozk}.
    
    \item The bulk theory we propose can also be considered as a gravity dual of a BCFT with corners \cite{Cardy:1989da, Imamura:2005zm, Geng:2021iyq}.
    In this duality, the corner on the boundary is in direct correspondence with the defect on the EOW branes. 
    
    \item The holographic entanglement entropy in our model exhibits three different phases. 
    This entanglement entropy can be seen as a toy model of a matter state prepared by cosmological spacetimes \cite{Chen:2020tes, Miyaji:2021lcq}.
    Interestingly, we have a phase that breaks the time reflection symmetry, so the corresponding entanglement entropy is closely related to the pseudo entropy \cite{Nakata:2020luh}.
    
    \item The defect enables us to construct a wormhole saddle that connects multiple AdS boundaries, analogous to the replica wormhole with EOW branes considered in \cite{Penington:2019kki}. 
    In our model, the wormhole saddle is allowed only when the BCFT is non-unitary, and this saddle is always subdominant compared to the factorized saddle without any wormholes. 
    
\end{itemize}

The organization of this paper is as follows. 
In section \ref{section:Gravitydual}, we present our refined AdS/BCFT model and its connection to BCFT with a corner.
In section \ref{section:Strip}, we study the spectrum of BCFT on a strip, confirming that it has a richer spectrum in comparison with the conventional AdS/BCFT model. 
In section \ref{section:Entropy}, we compute the holographic entanglement entropy and describe its three phases.
In section \ref{section:Replica}, we construct wormhole geometries using our model. 
In section \ref{section:discussion}, we conclude with discussions and future directions.

\section{Gravity Dual}
\label{section:Gravitydual}

In this section, we describe our proposed AdS/BCFT model in which the EOW branes are connected at a defect. The geometry with such intersecting EOW branes was previously considered in \cite{Geng:2021iyq}, and we will give an explicit model with action which realizes such geometry. We use our proposal to find the bulk dual of a BCFT with a corner.

In the original AdS/BCFT proposal\cite{Takayanagi:2011zk, Fujita:2011fp}, the holographic BCFT was dual to an asymptotically AdS spacetime $M$ with an EOW brane $\Sigma$ anchored to the BCFT boundary.
This EOW brane might contain matter fields and have a non-trivial gravitational action.
We take the bulk theory to be Einstein gravity 
\begin{equation}
    I_{\text{EH}} = -\frac{1}{16\pi G_N} \int_M \sqrt{g} \left(R-2\Lambda\right) -\frac{1}{8\pi G_N} \int_{N}\sqrt{h}K,
\end{equation}
where $N$ is the AdS boundary.

The simplest EOW brane action is given by the Gibbons-Hawking boundary term plus the tension term, 
\begin{equation}
    I_{\text{ETW}}
    =-\frac{1}{8\pi G_N}\int_{\Sigma}\sqrt{h}K+\frac{1}{8\pi G_N}\int_{\Sigma}\sqrt{h}T-\frac{1}{8\pi G_N}\int_{\Gamma_{\Sigma,N}}\sqrt{g_{\Gamma_{\Sigma,N}}}\left(\pi-\theta_{(\Sigma,N)}\right).
\end{equation}
Here $h_{ab}$ is the induced metric and $K_{ab}$ is the extrinsic curvature defined using the outgoing normal vector.
We also need to include the Hayward term for $\Gamma_{\Sigma,N}$, the corner between AdS boundary $N$ and the EOW brane $\Sigma$, with internal angle $\theta_{(\Sigma,N)}$.
This action leads to the equation of motion for the EOW brane
\begin{gather}
    K_{ab} = (K-T) h_{ab}.
\end{gather}

The boundary entropy $S_B$ in $2$d BCFT is defined in terms of the disk partition function via $Z_{\text{Disk}} =  e^{S_B}$ \cite{Affleck:1991tk}, which counts the number of degrees of freedom on the boundary. 
The tension $T$ of the EOW brane is related to the boundary entropy as
\begin{equation}
    S_B = \frac{c}{6} \atanh (\LR T),
\end{equation}
where $\LR$ is the AdS radius \cite{Azeyanagi:2007qj}.
In this simple model, there are no interactions and no intersections between EOW branes dual to distinct BCFT boundaries whereas for identical boundaries, the EOW branes can be smoothly connected.
This simple interaction leads to a simple value for the lowest eigenvalue \cite{Miyaji:2021ktr} that is highly non-generic, which we will discuss in section \ref{section:Strip}. 

\begin{figure}
    \centering
    \includegraphics[scale=0.6]{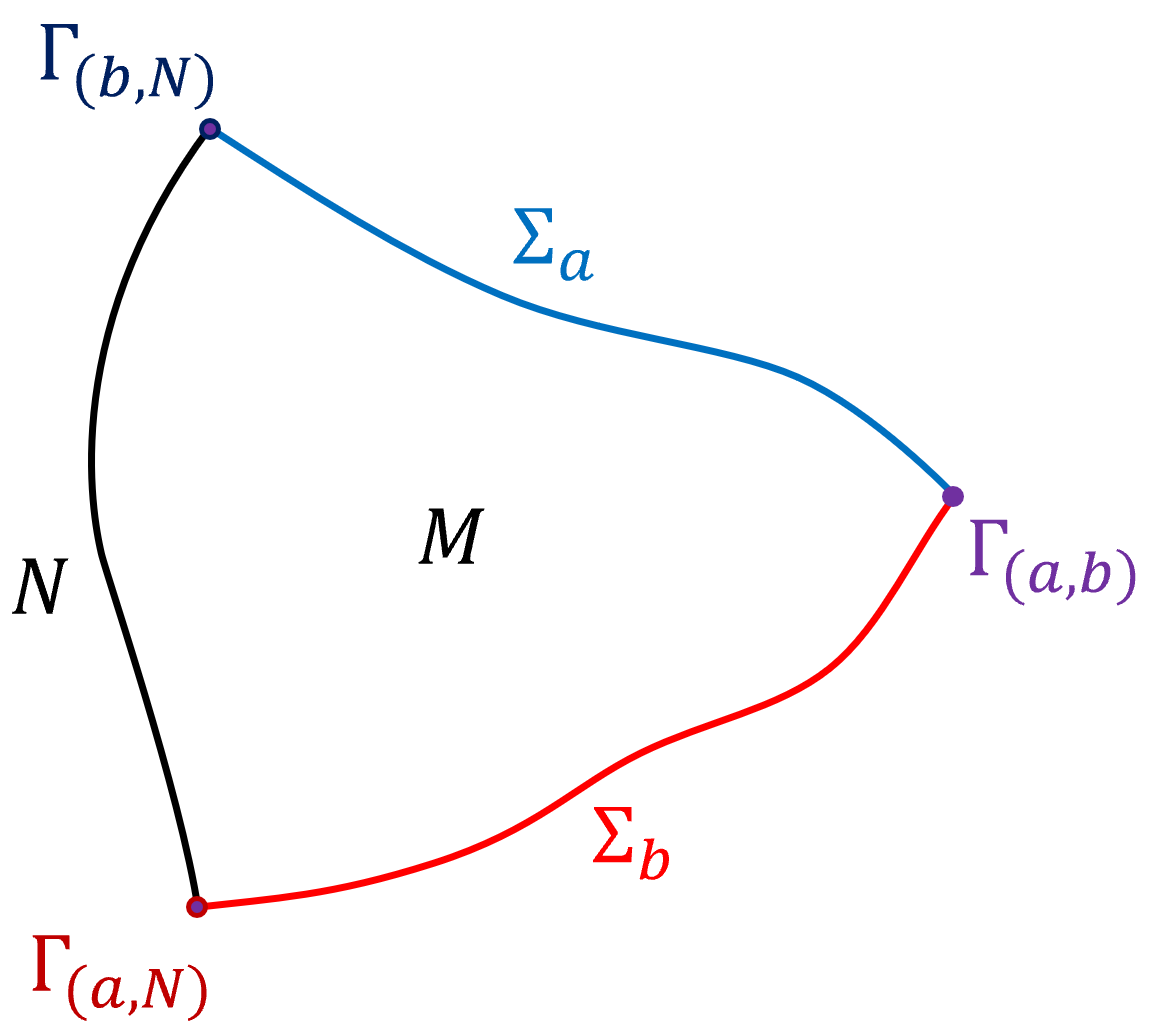}
    \caption{Sketch of our proposed gravity dual of BCFT.
    $N$ is the AdS boundary where the BCFT lives.
    $\Sigma_a$ and $\Sigma_b$ are the EOW branes, and $\Gamma_{(a,b)}$ is the defect connecting them.}
    \label{fig:schematic}
\end{figure}

The new ingredient introduced in this paper is such an interaction in the form of a defect that glues two distinct EOW branes. 
Since the dual spacetime terminates at some finite depth, the spectrum is no longer that simple, and indeed we have a new lowest eigenvalue as we show later.
The simplest action for this defect is given by
\begin{equation}\label{eq:Hayward}
    I_{\text{defect}}=
    -\frac{1}{8\pi G_N}\int_{\Gamma_{(a,b)}}\sqrt{g_{\Gamma_{(a,b)}}}\left(\theta_{0:(a,b)}-\theta_{(a,b)}\right).
\end{equation}
Here $\Gamma_{(a,b)}$ is the defect which glues two EOW branes $\Sigma_a$ and $\Sigma_b$, and $\theta_{(a,b)}$ is the internal angle of $M$ at $\Gamma_{(a,b)}$, see Fig.~\ref{fig:schematic}.
Also, $\theta_{0:(a,b)} - \pi$ can be regarded as the tension of the defect,
The action in (\ref{eq:Hayward}) is called Hayward term \cite{Hayward:1993my} when $\theta_{0:(a,b)} = \pi$, and has been considered in different AdS/BCFT contexts in \cite{Nozaki:2012qd, Takayanagi:2019tvn, Akal:2020wfl}.
This Hayward term can be obtained by taking a singular limit of the Gibbons-Hawking-York term where the boundary has a sharp corner.

The total action is the sum of these three contributions,
\begin{equation}
\label{eq:I_total}
    I_{\text{total}} = I_{\text{EH}} + I_{\text{ETW}} + I_{\text{defect}}.
\end{equation}
It is worth mentioning that this $I_{\text{total}}$ has UV divergences coming from the region near the AdS boundary.
To obtain a finite answer, these divergences need to be cured by introducing a short distance cutoff at the boundary and including appropriate counterterms to cancel the divergence.

The variation of $I_{\text{total}}$  is
\begin{multline}
    \delta I_{\text{total}} = -\frac{1}{16\pi G_N} \int_M \sqrt{g} \left( R_{\mu\nu} - \frac{1}{2}R g_{\mu\nu} + \Lambda g_{\mu\nu} \right) \delta g^{\mu\nu}
    \\
    - \sum_{i=a,b} \frac{1}{8\pi G_N} \int_{\Sigma_i} \sqrt{h} \left( K_{\alpha\beta} - K h_{\alpha\beta} + T_i h_{\alpha\beta} \right) \delta h^{\alpha\beta}
    - \frac{1}{8\pi G_N} \int_{N} \sqrt{h} \left( K_{\alpha\beta} - K h_{\alpha\beta} \right) \delta h^{\alpha\beta}
    \\
    -\frac{1}{8\pi G_N} \int_{\Gamma_{(a,b)}} \left( \theta_{0:(a,b)} - \theta_{(a,b)} \right) \delta\sqrt{g_{\Gamma_{(a,b)}}} 
    - \sum_{i=a,b} \frac{1}{8\pi G_N} \int_{\Gamma_{(i,N)}} \left( \pi - \theta_{(i,N)} \right) \delta\sqrt{g_{\Gamma_{(i,N)}}}.
\end{multline}

In AdS/BCFT, the metric can fluctuate freely on $\Sigma_i$, but not at AdS boundary $N$, therefore
\begin{equation}
    \begin{cases}
     \delta h^{\alpha\beta}_{N}=0 ,&\delta h^{\alpha\beta}_{\Sigma_{i}}~~\text{free},\\
     \delta\sqrt{g_{\Gamma_{(i,N)}}}=0,&
     \delta\sqrt{g_{\Gamma_{(a,b)}}}~~\text{free}.
    \end{cases}
\end{equation}
These boundary conditions at $\Sigma_i$ and $\Gamma_{(a,b)}$ result in the following equations of motion
\begin{gather}
    \label{eq:eom_KT}
    K_{\alpha\beta} = (K -T_i) h_{\alpha\beta}, \\
    \theta_{0:(a,b)} - \theta_{(a,b)} = 0.
    \label{eq:eom_theta0}
\end{gather}
These two equations determine the shapes of the EOW branes $\Sigma_i$ and of the corner $\Gamma_{(a,b)}$.
On the other hand, the above boundary conditions do not induce equations of motion at the AdS boundary and $\Gamma_{(a,N)}$ because of the Dirichlet boundary condition.

\subsection{Gravity Dual of BCFT with a Corner}

In this subsection, we examine the gravity dual of BCFT with a corner as a simple example.
This dual is given by a vacuum AdS spacetime with two EOW branes glued at a defect. 

\begin{figure}
    \centering
    \includegraphics[scale=0.6]{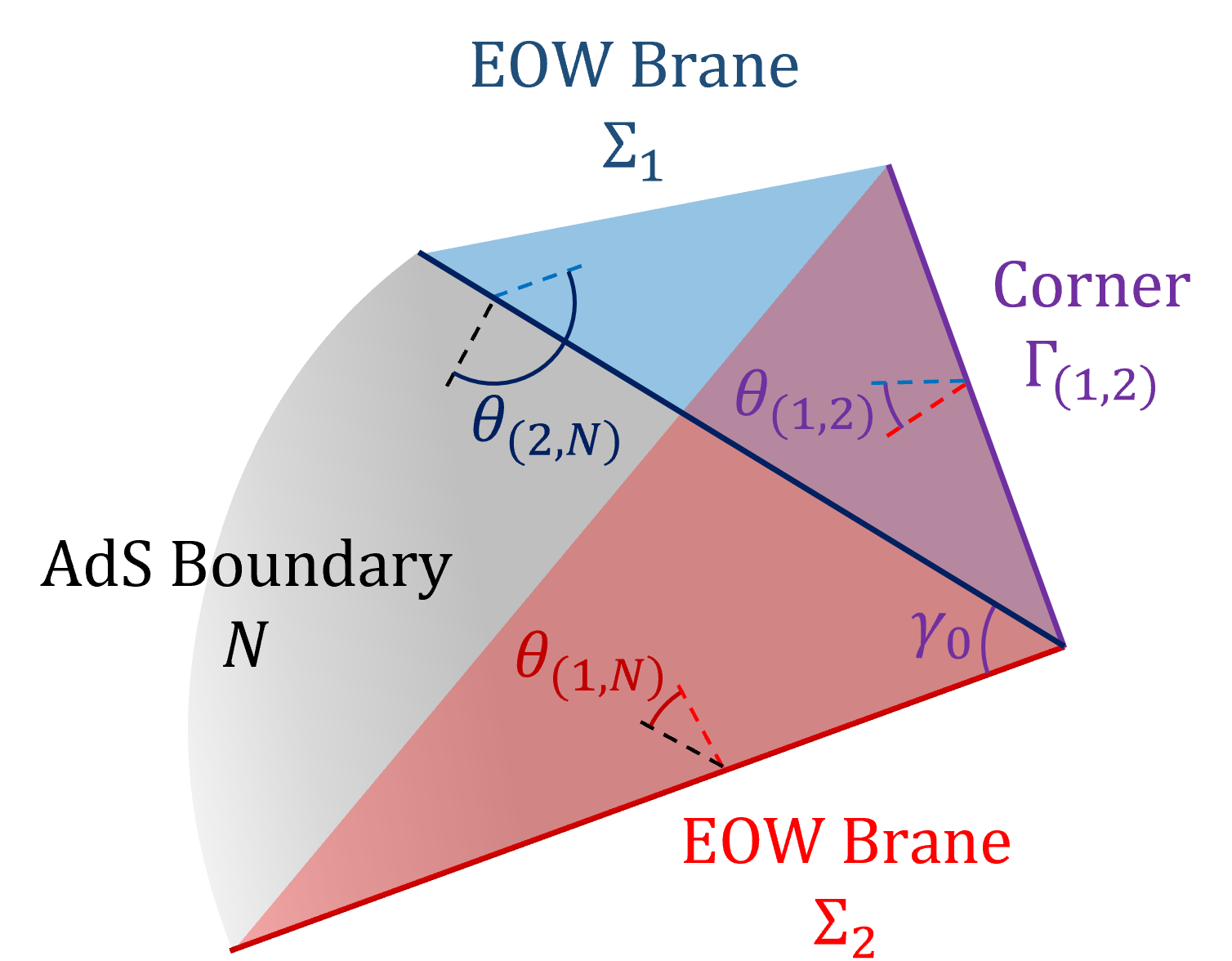}
    \caption{Two EOW branes embedded in Poincare AdS. 
    They intersect in the bulk at an internal angle $\theta_{(1,2)}$ and on the boundary at an internal angle $\gamma_0$. 
    The bulk region is dual to BCFT on the cornered region $N = D$. }
    \label{fig:poincare}
\end{figure}

Consider the Poincare background for vacuum AdS$_3$
\begin{equation}
    \d s^2=\frac{\LR^2}{z^2}(\d x^2 + \d y^2 + \d z^2)
\end{equation}
with $\LR$ being the AdS length.
We restrict to the region 
\begin{equation}
    D := \{ u=r e^{\i\theta} \in \mathbb{C} | r \geq 0, ~0 \leq \theta \leq \gamma_0 \}
\end{equation}
on the AdS boundary\footnote{The region $D$ can be mapped to the upper half plane via a singular conformal transformation, and standard BCFT techniques can be applied \cite{Geng:2021iyq}. See also \cite{Cardy:1989da}.} with $u := x + \i y$.
Since this region has a corner at the origin, the bulk geometry is dual to BCFT with a corner.
The bulk spacetime  we consider is a spacetime bounded by two EOW branes anchored to $\partial D$. 
The equation of motion for the EOW brane fixes its shape to be a plane in $(x,y,z)$ coordinates.
Fig.~\ref{fig:poincare} shows the bulk geometry with these two intersecting EOW branes.
Since the bulk metric is the Poincare metric, the Fefferman-Graham expansion tells us that the stress tensor $\langle T_{uu}\rangle$ vanishes away from the boundaries.

Let us first assume that the two EOW branes have equal tensions, so the boundary entropies for the two boundaries are identical.
The dimensionless tension $\TT_a = \LR T_a$ of the EOW brane is related to the internal angle $\theta_{(a,N)}$ between the brane and the AdS boundary as
\begin{equation}
    \TT_a = - \cos \theta_{(a,N)}.
\end{equation}
Note that the allowed values of the tension are given by $|\TT|\leq 1$.

The internal angle $\gamma_0$ at AdS boundary and the internal angle $\theta_{(1,2)}$ between EOW branes are related as
\begin{equation}
    \gamma_0 = 
    \begin{cases}
     \acos \left( \frac{\cos \theta_{(1,2)} + \TT^2}{1-\TT^2} \right) &\quad 2\asin |\TT|<\theta_{(1,2)}<\pi\\
    2\pi -  \acos \left( \frac{\cos \theta_{(1,2)} + \TT^2}{1-\TT^2} \right) &\quad \pi<\theta_{(1,2)}<2\pi - 2\asin |\TT|
    \end{cases}.
\end{equation}
Since the configuration of the EOW brane is a plane in the $(x,y,z)$ coordinates, $\theta_{(1,2)}$ is simply the angle of intersection of two planes. 

For arbitrary tensions $\TT_1$ and $\TT_2$, we have
\begin{equation}\label{eq:gamma0}
    \gamma_0 = 
    \begin{cases}
    \acos \left( \frac{\cos \theta_{(1,2)} + \TT_1\TT_2}{\sqrt{(1-\TT_1^2)(1-\TT_2^2)}} \right),\\
    \quad\quad\quad\quad \acos \left(\text{\footnotesize ${\sqrt{(1-\TT_1^2)(1-\TT_2^2)} - \TT_1 \TT_2}$ } \right) <\theta_{(1,2)}< \pi - \acos \left(\text{\footnotesize ${\sqrt{(1-\TT_1^2)(1-\TT_2^2)} + \TT_1 \TT_2}$}\right), \\
    2\pi - \acos \left( \frac{\cos \theta_{(1,2)} + \TT_1\TT_2}{\sqrt{(1-\TT_1^2)(1-\TT_2^2)}} \right),\\
    \quad\quad\quad\quad \pi +\acos  \left(\text{\footnotesize ${\sqrt{(1-\TT_1^2)(1-\TT_2^2)} + \TT_1 \TT_2}$ } \right) <\theta_{(1,2)}< 2\pi - \acos \left(\text{\footnotesize ${\sqrt{(1-\TT_1^2)(1-\TT_2^2)} - \TT_1 \TT_2}$} \right).
    \end{cases}
\end{equation}
In the following sections, we will use this geometry to construct the gravity dual of BCFT on a strip.

\section{BCFT on a Strip} \label{section:Strip}

In this section, we construct the gravity dual of BCFT on an infinite Euclidean strip by using our refined model for AdS/BCFT.
We suppose that conformal boundary conditions $a$ and $b$ have been imposed on the two boundaries of this strip.

Let us first review the conventional AdS/BCFT model which does not have any defects connecting the EOW branes.
When the two boundary conditions are identical, the dual is given by thermal AdS with the EOW branes being connected without any defect (if the two boundaries are sufficiently close to each other) \cite{Takayanagi:2011zk, Fujita:2011fp, Miyaji:2021ktr}. 
In this case, the lowest eigenvalue was found to be $ E_{(a,b)}^{ \text{BCFT} } = - \frac{\pi c}{24\Delta x}$ where $\Delta x$ is the width of the strip.
This spectrum corresponds to the conformal dimension of the boundary condition changing operator being $h_{(a,b)}^{\text{bcc}} = 0 $.

When the boundary entropies of two boundary conditions are different, the dual geometry is given by two disconnected EOW branes in the Poincare background. 
As the result, the lowest eigenvalue for the BCFT was found to be $E_{(a,b)}^{\text{BCFT}}=0$, which corresponds to $h_{(a,b)}^{\text{bcc}} = \frac{c}{24} $. 
These two results imply that there is no operator with conformal dimension between $h_{(a,b)}^{\text{bcc}} = 0$ and $h_{(a,b)}^{\text{bcc}} = \frac{c}{24}$ in the conventional AdS/BCFT model.

In the following subsections, we generalize this simple spectrum by introducing a defect that can connect the EOW branes, even if they have different tensions, or equivalently, different boundary entropies.
This will allow us to obtain any value for the lowest operator dimension between $h_{(a,b)}^{\text{bcc}} = 0$ and $h_{(a,b)}^{\text{bcc}} = \frac{c}{24}$. 

\subsection{Bulk Geometry}

We are interested in constructing the gravity dual of an infinite Euclidean strip of width $\Delta x$.
Let us start by considering the thermal AdS$_3$ geometry with two EOW branes connected through a defect. 
The metric of the thermal AdS$_3$ without any branes is
\begin{equation}
\label{eq:metric}
    \d s^2 = \LR^2 \left( \frac{\d\tau^2}{z_0^2 \chi^2} + \frac{\d \chi^2}{h(\chi) \chi^2} + \frac{ h(\chi)}{\chi^2} \d \phi^2 \right).
\end{equation}
Here $h(\chi) = (1-\chi^2)$, with $0 < \chi \leq 1$. 
We denote the periodicity of the Euclidean time $\tau$ as $T_{\mathrm{BCFT}}^{-1} = 2 \pi z_0 \chi_H$ which is assumed to be large compared to $z_0$ so that the thermal AdS saddle is dominant compared to Euclidean black hole.
The $\phi$ coordinate has period $2 \pi$ so that there is no conical defect at $\chi = 1$. 
Note that $z_0$ sets the length scale for the BCFT.

Let us first embed a single EOW brane without any defects in this geometry.
Since the brane satisfies the equation of motion in \eqref{eq:eom_KT}, its profile is
\begin{equation}
\label{eq:profile}
    \phi(\chi) = \phi(0) \pm \atan  \text{\small $ \frac{\TT \chi}{ \sqrt{h(\chi) - \TT^2}} $} ,
\end{equation}
in terms of the dimensionless tension $\TT$. 
As explained in appendix \ref{appendix:conformal_transformation}, the bulk coordinate transformation in \eqref{eq:bulktransformation} can be used to show that this configuration is equivalent to EOW brane with tension $\TT$ anchored to a line in Poincare AdS. 
The deepest point that the EOW brane reaches before turning back towards the AdS boundary is
\begin{equation}
    \chi = \chi_0(\TT) := \sqrt{1-\TT^2}.
\end{equation}

Next we consider two EOW branes $\Sigma_1$ and $\Sigma_2$, with dimensionless tensions $\TT_1$ and $\TT_2$, respectively.
We assume that these tensions satisfy $|\TT_2| \leq |\TT_1|$, without loss of generality. 
The branes are anchored to the AdS boundary at $\Gamma_{1,N} = \{ \chi = 0,\, \phi = \pi - \alpha \}$ and $\Gamma_{2,N} = \{ \chi = 0,\, \phi = \pi + \alpha \}$ respectively, with some $\alpha \in (0,\pi)$.
As mentioned earlier, $M$ denotes the bulk region between these branes.
Its boundary $\partial M$ consists of the two EOW branes $\Sigma_1$ and $\Sigma_2$, along with a region on the AdS boundary, $N = \{ \chi = 0, \pi - \alpha \leq \phi \leq \pi + \alpha \}$.
From the perspective of the AdS boundary, the angular size of $N$ is $\Delta\phi = 2\alpha$.
The width of the strip is $\Delta x = 2\alpha z_0$.

We want to calculate the angle $\theta_{(1,2)}$ at which the two EOW branes intersect in the bulk. 
Depending on the value of $\alpha$, the bulk geometries are qualitatively different, so we need to treat the various cases separately. 

\begin{figure}
    \centering
    \includegraphics[scale=0.6]{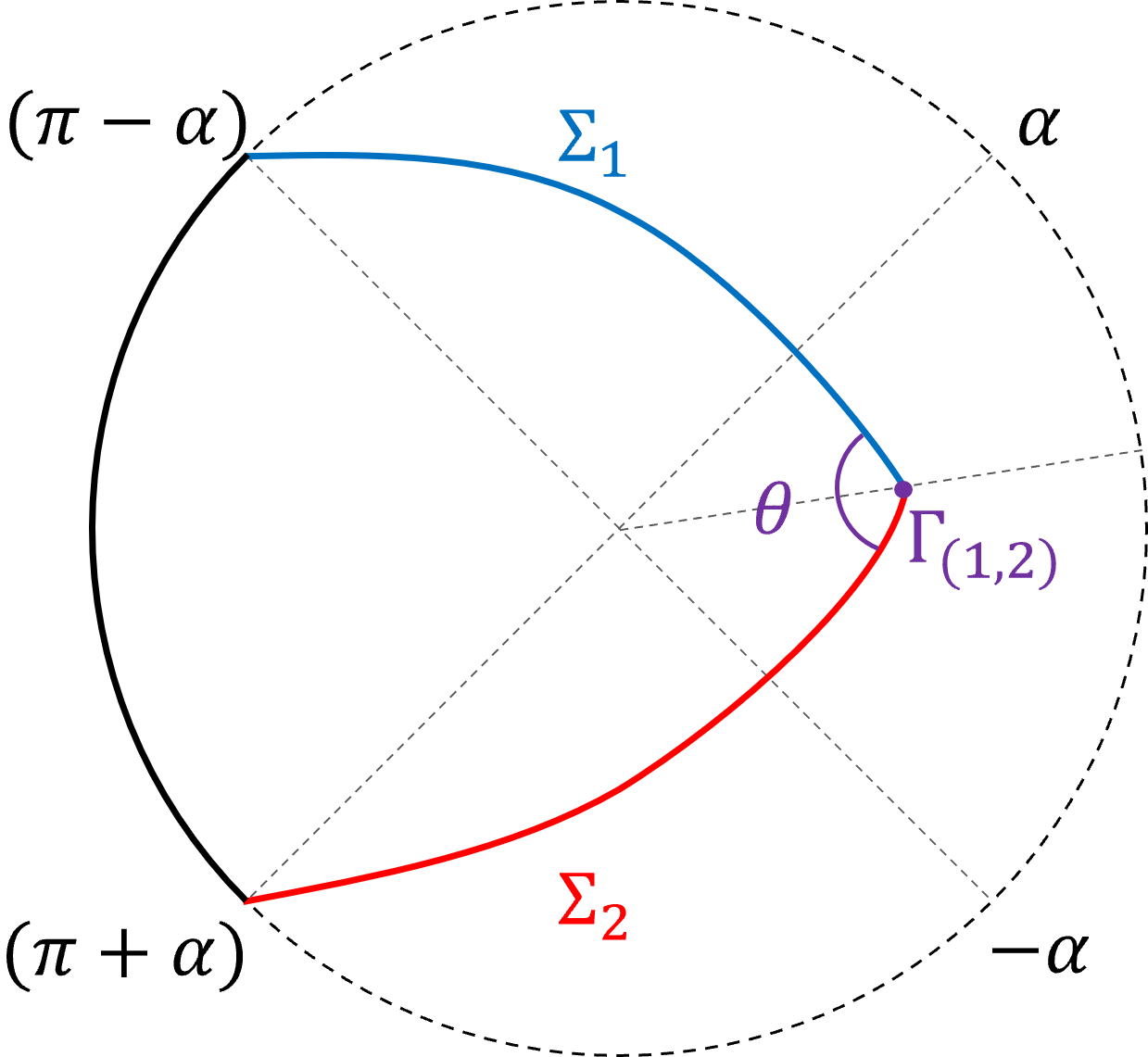}
    \caption{A constant $\tau$ slice of thermal AdS with two EOW branes, $\Sigma_1$ and $\Sigma_2$, meeting at the corner $\Gamma_{(1,2)}$. }
    \label{fig:I_bulk_eg}
\end{figure}

As an illustration, let us consider the case where $\TT_1, \TT_2>0$ and $\alpha$ is small, in particular, $0 \leq \alpha \leq \frac{\pi}{4} + \frac{1}{2} \atan \frac{ \TT_2 \sqrt{1-\TT_1^2} }{ \sqrt{\TT_1^2-\TT_2^2} }$.
We will state the results for the other cases momentarily. 
A constant $\tau$ slice of the bulk geometry, in this case, is shown in  Fig.~\ref{fig:I_bulk_eg}.
The profiles of the EOW branes in the intersection region are 
\begin{equation}
    \phi_1(\chi) = -\alpha + \atan \text{\small $ \frac{\TT_1 \chi}{\sqrt{h(\chi) - \TT_1^2}} $} ,
    \quad\quad
    \phi_2(\chi) = \alpha - \atan \text{\small $\frac{\TT_2 \chi}{\sqrt{h(\chi) - \TT_2^2}} $} .
\end{equation}

To obtain the intersection point $\chi_*$, we set $\phi_1(\chi_*) = \phi_2(\chi_*)$ to get an equation for $\chi_*$
\begin{gather}
\label{eq:chi_s_1}
    \atan \text{\small $ \frac{\TT_1 \chi_*}{\sqrt{h(\chi_*) - \TT_1^2}} $} + \atan \text{\small $ \frac{\TT_2 \chi_*}{\sqrt{h(\chi_*) - \TT_2^2}} $} = 2 \alpha.
\end{gather}
This $\chi_*$ increases monotonically with $\alpha$.
The upper bound on $\alpha$ for this case is attained when $\chi_*$ equals the maximum value $\chi_{0}(\TT_1) = \sqrt{1-\TT_1^2}$
\footnote{This maximal value corresponds to the deepest point on $\Sigma_1$ because we have assumed that $|\TT_1| > |\TT_2|$, which implies that $\chi_{0}(\TT_1) < \chi_{0}(\TT_2)$.}, and the explicit value is $\alpha = \frac{\pi}{4} + \frac{1}{2} \atan \frac{ \TT_2 \sqrt{1-\TT_1^2} }{ \sqrt{\TT_1^2-\TT_2^2} }$ as stated above.
The angle of intersection between the two EOW branes is
\begin{equation}
\label{eq:theta_n1n2_1}
    \theta_{(1,2)} = \acos \text{\small $ \frac{\sqrt{ (h(\chi_*)-\TT_1^2) (h(\chi_*)-\TT_1^2) } - \TT_1\TT_2 }{ h(\chi_*) } $} .
\end{equation}

Now we state the result for general $\alpha$.
Assuming that the tensions satisfy the condition $|\TT_2| \leq |\TT_1|$, the equation to obtain the intersection point is
\begin{equation}
    \begin{cases}
    \Big| \atan \frac{\TT_1 \chi_* }{ \sqrt{h(\chi_*) - \TT_1^2}} + \atan \frac{\TT_2 \chi_*}{\sqrt{h(\chi_*) - \TT_2^2}} \Big| = 2 \alpha ,
    & 0<\alpha< \frac{\pi}{4} +  \frac{\sgn(\TT_1)}{2} \atan \frac{\TT_2 \sqrt{1-\TT_1^2}}{\sqrt{\TT_1^2-\TT_2^2}} ,\\
    \Big| \atan \frac{\TT_1 \chi_*}{\sqrt{h(\chi_*) - \TT_1^2}} - \atan \frac{\TT_2 \chi_*}{\sqrt{h(\chi_*) - \TT_2^2}}\Big| = \pi - 2 \alpha ,
    & \frac{\pi}{4} + \frac{\sgn(\TT_1)}{2} \atan \frac{\TT_2 \sqrt{1-\TT_1^2} }{ \sqrt{\TT_1^2-\TT_2^2} } < \alpha < \frac{\pi}{2} ,\\
    \Big| \atan \frac{\TT_1 \chi_*}{\sqrt{h(\chi_*) - \TT_1^2}} - \atan \frac{\TT_2 \chi_*}{\sqrt{h(\chi_*) - \TT_2^2}} \Big| = 2 \alpha - \pi ,
    & \frac{\pi}{2} < \alpha < \frac{3\pi}{4} - \frac{\sgn(\TT_1)}{2} \atan \frac{ \TT_2 \sqrt{1-\TT_1^2}}{\sqrt{\TT_1^2-\TT_2^2}} ,\\
    \Big| \atan \frac{\TT_1 \chi_*}{\sqrt{h(\chi_*) - \TT_1^2}} + \atan \frac{\TT_2 \chi_*}{\sqrt{h(\chi_*) - \TT_2^2}}\Big| = 2\pi - 2 \alpha ,
    & \frac{3\pi}{4} - \frac{\sgn(\TT_1)}{2} \atan \frac{ \TT_2 \sqrt{1-\TT_1^2} }{ \sqrt{\TT_1^2-\TT_2^2}} < \alpha < \pi ,
    \end{cases}
\end{equation}
and the angle of intersection is
\begin{equation}
    \theta_{(1,2)} = 
    \begin{cases}
    \acos \left( \frac{\sqrt{ (h(\chi_*)-\TT_1^2) (h(\chi_*)-\TT_2^2) } - \TT_1\TT_2 }{ h(\chi_*) } \right),
    & 0<\alpha< \frac{\pi}{4} +  \frac{\sgn(\TT_1)}{2} \atan \frac{\TT_2 \sqrt{1-\TT_1^2}}{\sqrt{\TT_1^2-\TT_2^2}}, \\
    \acos \left( - \frac{\sqrt{ (h(\chi_*)-\TT_1^2) (h(\chi_*)-\TT_2^2) } + \TT_1\TT_2 }{ h(\chi_*) } \right),
    & \frac{\pi}{4} + \frac{\sgn(\TT_1)}{2} \atan \frac{\TT_2 \sqrt{1-\TT_1^2} }{ \sqrt{\TT_1^2-\TT_2^2} } < \alpha < \frac{\pi}{2}, \\
    2 \pi - \acos \left( - \frac{\sqrt{ (h(\chi_*)-\TT_1^2) (h(\chi_*)-\TT_2^2) } + \TT_1\TT_2 }{ h(\chi_*) } \right),
    & \frac{\pi}{2} < \alpha < \frac{3\pi}{4} - \frac{\sgn(\TT_1)}{2} \atan \frac{\TT_2 \sqrt{1-\TT_1^2}}{\sqrt{\TT_1^2-\TT_2^2}}, \\
    2\pi - \acos \left( \frac{\sqrt{ (h(\chi_*)-\TT_1^2) (h(\chi_*)-\TT_2^2) } - \TT_1\TT_2 }{ h(\chi_*) } \right),
    & \frac{3\pi}{4} - \frac{\sgn(\TT_1)}{2} \atan \frac{\TT_2 \sqrt{1-\TT_1^2} }{ \sqrt{\TT_1^2-\TT_2^2}} < \alpha < \pi .
    \end{cases}
\label{eq:theta_1234}
\end{equation}
Here, $\sgn(x) := \frac{x}{|x|}$ is the sign function.
Note that these results are valid even when one or both tensions are negative.

The equation of motion \eqref{eq:eom_theta0} fixes the angle of intersection to be $\theta_{(1,2)} = \theta_0$.
This determines the value of $\alpha$, and hence, the geometry in terms of $\theta_0$ and the tensions of EOW branes.
Indeed, substituting $\theta_{(1,2)} = \theta_0$ in \eqref{eq:theta_1234}, we get
\begin{equation}
\label{eq:chi}
    \chi_* = \sqrt{ 1 - \text{\small $ \frac{ \TT_1^2 + \TT_2^2 + 2\TT_1\TT_2 \cos\theta_0 }{ \sin^2\theta_0 } $} }.
\end{equation}
Also, the value of $\alpha$ that corresponds to this $\theta_0$ is
\begin{equation}
\label{eq:alpha0}
    \alpha_0 = 
    \begin{cases}
    \frac{1}{2} \acos \left( \frac{\cos \theta_0 + \TT_1\TT_2}{\sqrt{(1-\TT_1^2)(1-\TT_2^2)}} \right),\\
    \quad\quad\quad\quad \acos \left(\text{\footnotesize ${\sqrt{(1-\TT_1^2)(1-\TT_2^2)} - \TT_1 \TT_2}$ } \right) <\theta_0< \pi - \acos \left(\text{\footnotesize ${\sqrt{(1-\TT_1^2)(1-\TT_2^2)} + \TT_1 \TT_2}$}\right), \\
    \pi - \frac{1}{2} \acos \left( \frac{\cos \theta_0 + \TT_1\TT_2}{\sqrt{(1-\TT_1^2)(1-\TT_2^2)}} \right),\\
    \quad\quad\quad\quad  \pi + \acos\left(\text{\footnotesize ${\sqrt{(1-\TT_1^2)(1-\TT_2^2)} + \TT_1 \TT_2}$ } \right) <\theta_0< 2\pi - \acos \left(\text{\footnotesize ${\sqrt{(1-\TT_1^2)(1-\TT_2^2)} - \TT_1 \TT_2}$} \right).
    \end{cases}
\end{equation}
Note that \eqref{eq:chi} and \eqref{eq:alpha0} are symmetric in $\TT_1$ and $\TT_2$, therefore the formula applies even when $|\TT_1| < |\TT_2|$. 
Moreover, this result for $\alpha_0$ is identical to \eqref{eq:gamma0} if we identify $\gamma_0$ with $2\alpha_0$.
In fact, the two bulk geometries can be identified by a bulk coordinate transformation which maps the infinite strip $N$ to the cornered region $D$, as explained in appendix \ref{appendix:conformal_transformation}.

\begin{figure}
    \centering
\begin{subfigure}[b]{0.32\textwidth}
    \centering
    \includegraphics[width=\textwidth]{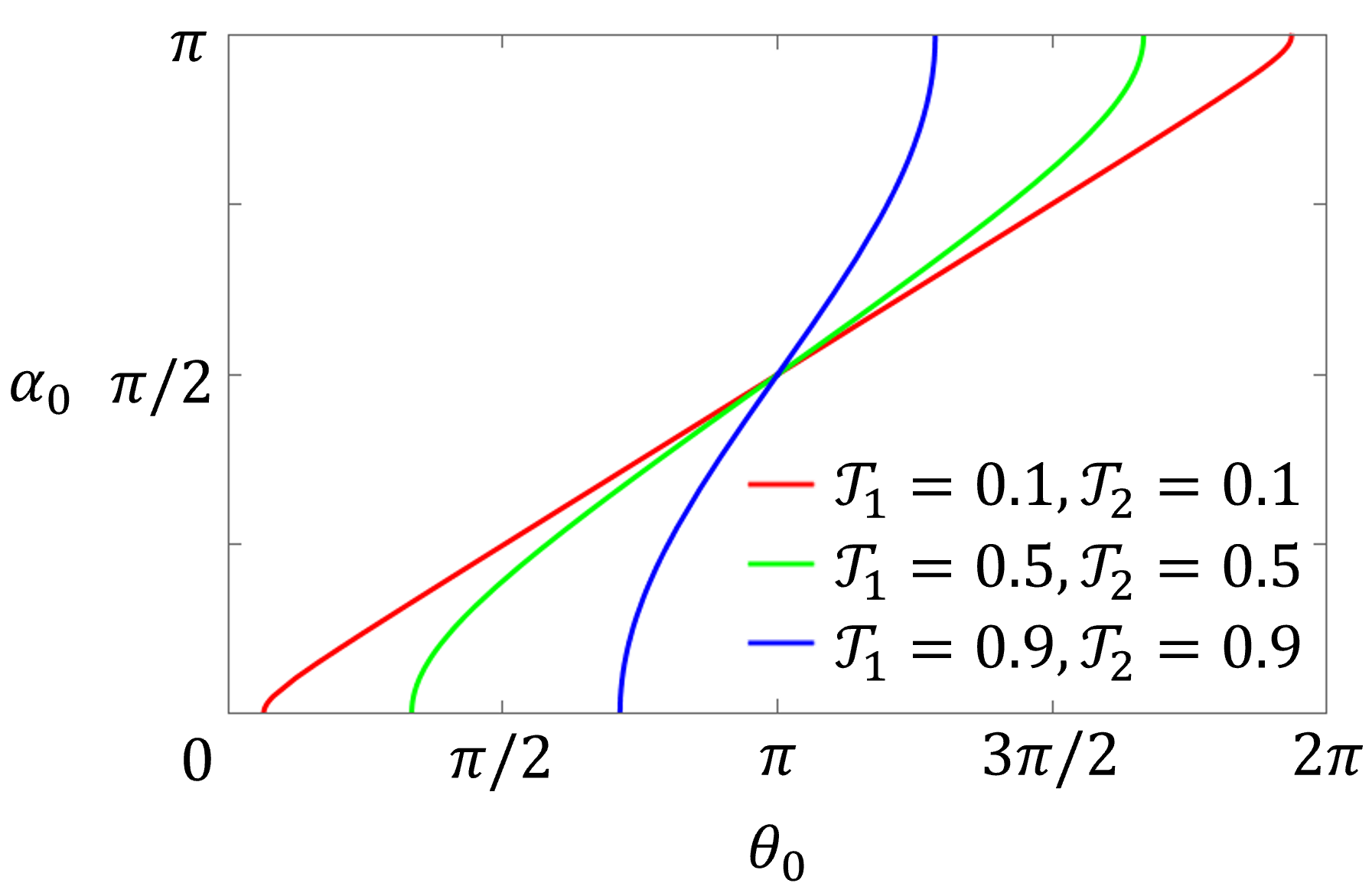}
    \caption{Equal tension $\TT_1=\TT_2$}
    \label{fig:at3}
\end{subfigure}
\hfill
\begin{subfigure}[b]{0.32\textwidth}
    \centering
    \includegraphics[width=\textwidth]{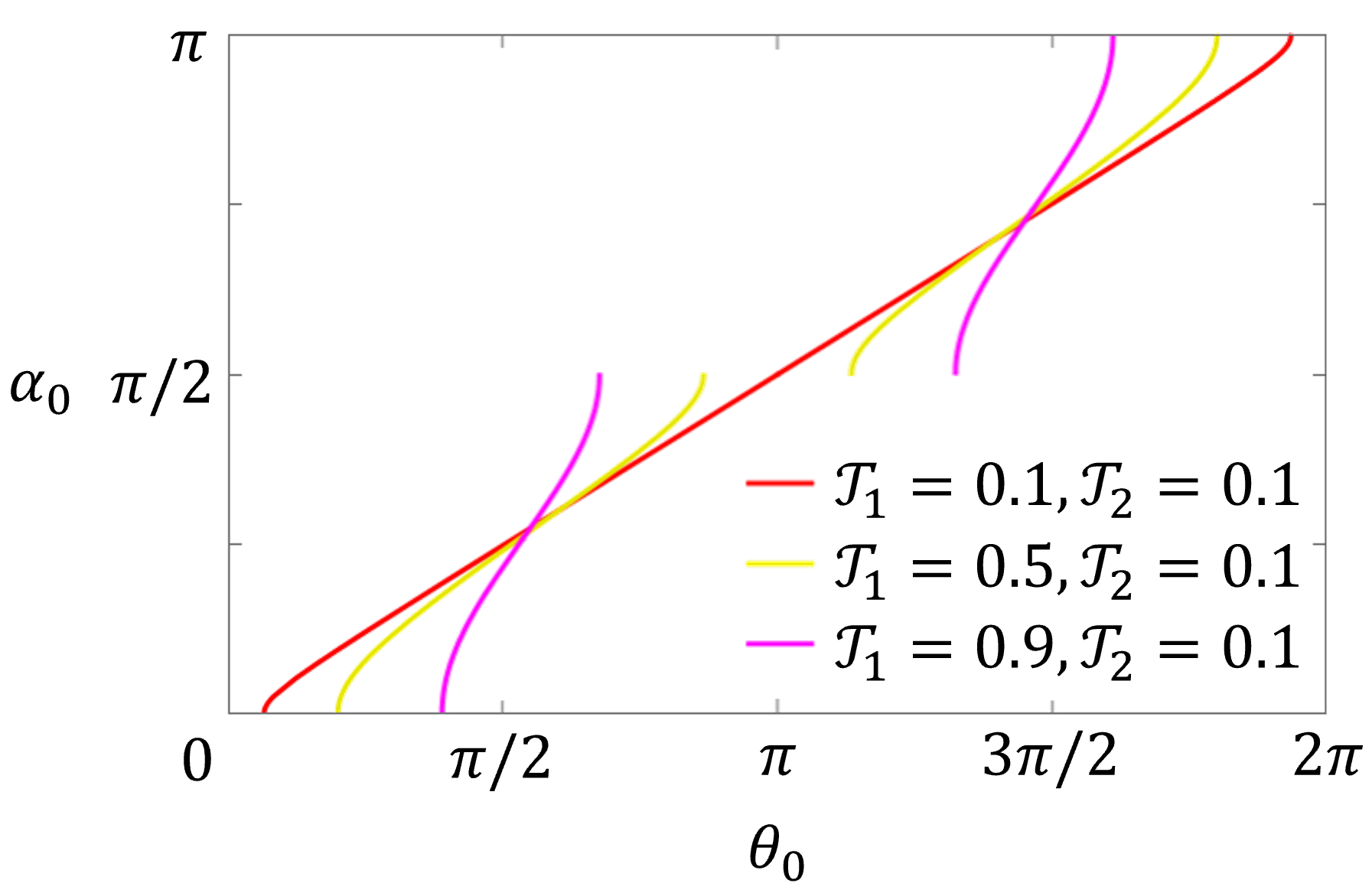}
    \caption{Fixed $\TT_2$}
    \label{fig:at1}
\end{subfigure}
\hfill
\begin{subfigure}[b]{0.32\textwidth}
    \centering
    \includegraphics[width=\textwidth]{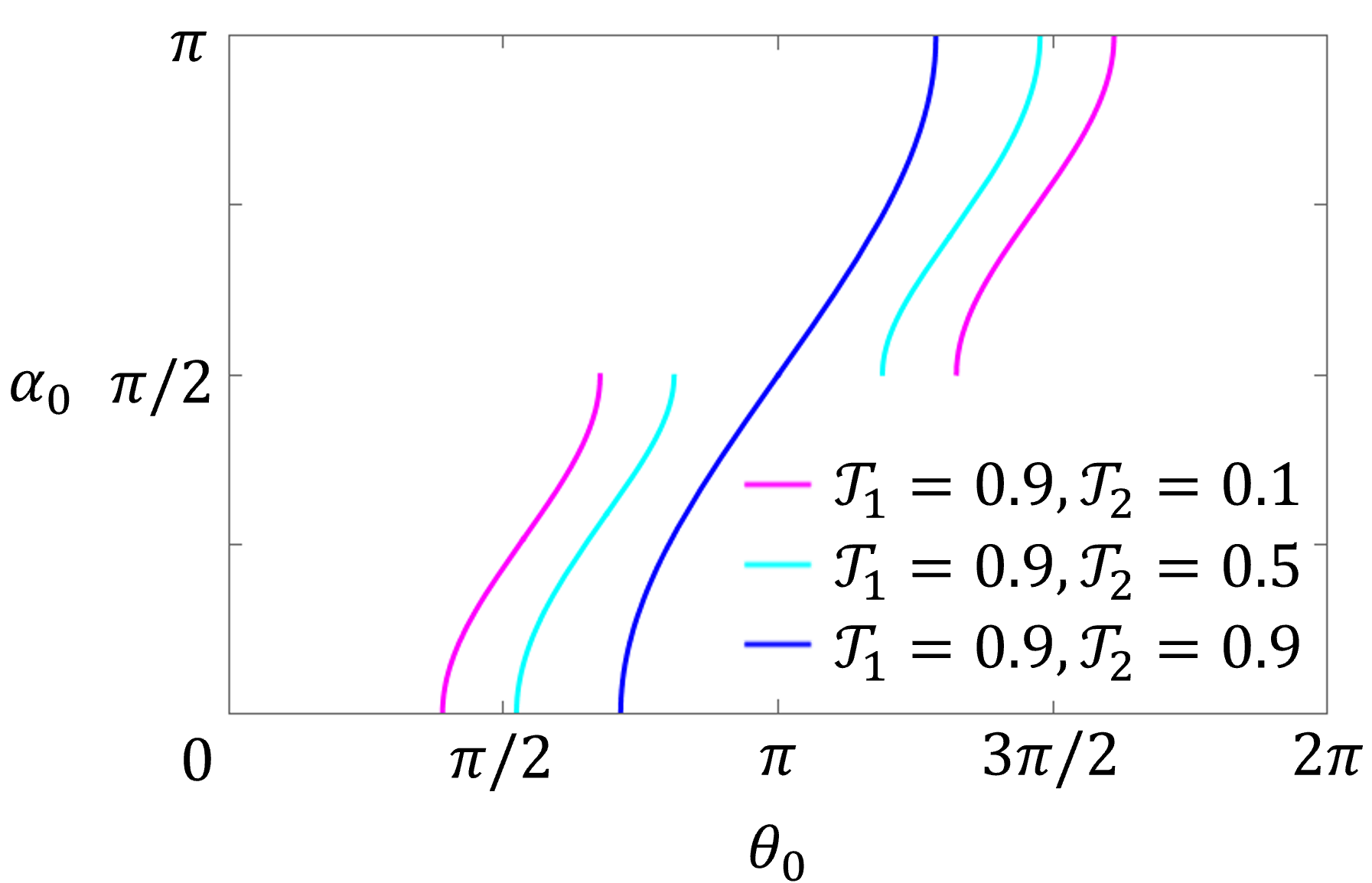}
    \caption{Fixed $\TT_1$}
    \label{fig:at2}
\end{subfigure}

    \caption{Plot of $\alpha_0$ as a function of $\theta_0$ for various values of tensions.}
    \label{fig:alpha_theta}
\end{figure}

Fig.~\ref{fig:alpha_theta} shows $\alpha_0$ as we vary $\theta_0$ for various values of the tensions based on \eqref{eq:alpha0}.
When the two tensions are equal $\TT_1 = \TT_2$, then $\theta_0$ is a continuous function of $\alpha_0$.
In this case, \eqref{eq:chi} and \eqref{eq:alpha0} simplify to
\begin{equation}
    \chi_* = \sqrt{1 - \text{\small $ \frac{\TT^2}{\sin^2(\theta_0/2)} $} },
\end{equation}
and
\begin{equation}
\label{eq:equaltensionangle}
    \alpha_0 = 
    \begin{cases}
    \frac{1}{2} \acos \left( \frac{\cos \theta_0 + \TT^2}{1-\TT^2} \right) , &\quad 2\asin |\TT| < \theta_0 \leq \pi ,\\
    \pi - \frac{1}{2} \acos \left( \frac{\cos \theta_0 + \TT^2}{1-\TT^2} \right) , &\quad \pi \leq \theta_0 < 2\pi - 2\asin |\TT| .
    \end{cases}
\end{equation}

If we also set $\theta_0 = \pi$, then we get $\alpha_0 = \frac{\pi}{2}$.
The dual bulk geometry has a smooth EOW brane geometry without any defect, which is the same as the conventional AdS/BCFT model.
Conversely, we find that $\theta_0 = \pi$ is possible only if $\TT_1 = \TT_2$.

When $\TT_1 \neq \TT_2$, the curve for $\theta_0$ has a discontinuity at $\alpha_0 = \pi/2$.
While all other values for $\alpha_0$ are valid, $\alpha_0 = \pi/2$ is disallowed.
This is because as we send $\theta_0$ to its corresponding value, the defect approaches the boundary and one of the EOW branes disappears.
For $\alpha_0 = \pi/2$, we only have one EOW brane and no defect, so this value needs to be excluded.

\vspace{0.5\baselineskip}
\noindent
\textbf{Conical defect in the bulk.}
So far, the geometries we have considered have a defect only on the EOW brane. 
We can construct more general geometries by including a bulk conical defect as well.
First, we consider the metric \eqref{eq:metric}, and embed the EOW branes whose profiles are given by \eqref{eq:profile} with internal angle $\theta_{(1,2)}$. 
A bulk conical defect located at $\chi=1$ with a deficit angle of $\Delta\phi$  can be introduced by identifying $\phi = \phi_0$ and  $\phi = \phi_0 + \Delta\phi$ in the geometry \eqref{eq:metric}. 
Here, $\phi_0$ is chosen such that the region given by $\phi \in [\phi_0, \phi_0 + \Delta\phi_0]$ is contained within the spacetime $M$ which is bounded by EOW branes. 
This is possible if and only if
\begin{equation}
    \TT_1, ~ \TT_2 > 0,
    \quad\quad\quad\quad
    2\alpha_0 \geq \Delta\phi.
\end{equation}
The first condition $\TT_1, ~ \TT_2 > 0$ is required for $\chi = 1$ to be contained in the geometry, and the second condition prevents self-intersection of the EOW branes \cite{Geng:2021iyq}. 
For more recent analysis, see \cite{Kawamoto:2022etl, Kusuki:2021gpt, Kusuki:2022wns, Kusuki:2022ozk, Numasawa:2022cni}.

\subsection{Euclidean Action}

In this subsection, we compute the Euclidean action for the bulk geometry dual to the strip.
This action is directly related to the lowest eigenvalue of the BCFT as we shall see in the next subsection.

First, consider the case with no conical defect in the bulk.
After an explicit computation, we find that the total action given by \eqref{eq:I_total} is
\begin{equation}
    I_{\text{total}} = \frac{\LR \chi_H}{2G_N} \left[ -\frac{\alpha_0}{\epsilon^2} + \frac{1}{2\epsilon} \sum_{a=1}^{2} \left( \frac{\TT_a}{ \sqrt{1-\TT_a^2} } - \acos\TT_a \right) \right],
\end{equation}
where we have introduced the short-distance cutoff $\epsilon$ at the AdS boundary to regulate the UV divergences.

To cancel these divergences, we need to include boundary counterterms.
These counterterms must be covariant and local, and are chosen to cancel the divergences.
In our case, 
\begin{equation}
\begin{split}
    I_{\text{ct}} &= \frac{1}{8\pi G_N \LR}\int_{N}\sqrt{h_{\partial M}} + \sum_{a} \frac{\acos\TT_a}{8\pi G_N} \int_{\Gamma_{(a,N)}} \sqrt{ g_{\Gamma_{(a,N)}} } \\
    &= \frac{\LR\chi_H}{2G_N} \left[ \alpha_0 \left( \frac{1}{\epsilon^2} - \frac{1}{2} \right) - \frac{1}{2\epsilon} \sum_{a=1}^{2} \frac{\TT_a}{ \sqrt{1-\TT_a^2} } \right] + \frac{\LR\chi_H}{4 G_N \epsilon} \sum_{a=1}^{2} \acos\TT_a.
\end{split}
\end{equation}
On including these counterterms, we obtain the Euclidean action
\begin{equation}
\label{eq:IE}
    I_E = - \frac{\LR \chi_H}{4G_N} \alpha_0 = -\frac{c}{6\pi\xi} \alpha_0^2.
\end{equation}
Here we have defined the aspect ratio $\xi := \Delta x \cdot T_{\mathrm{BCFT}}$ in order to have scale-invariant expressions. 
We have also used that the width of the strip is $\Delta x = 2 \alpha_0 z_0$ and the central charge is $c = \frac{3\LR}{2G_N}$.

As an example, let us consider the case $\TT_1 = \TT_2$ and $\alpha_0 = \frac{\pi}{2}$, for which the EOW branes do not have a defect.
Then the action is given by
\begin{equation*}
    I_E^{\mathrm{con}} = -\frac{\pi c}{24\xi},
\end{equation*}
where the superscript denotes that this is the answer in the connected phase of the conventional AdS/BCFT model. 

Let us generalize the above result to geometries with bulk conical defects, assuming $\TT_1, ~ \TT_2 > 0$.
The action can be obtained by subtracting off the contribution corresponding to the bulk portion that gets removed due to the conical defect. 
The result is 
\begin{equation}
\label{eq:IEconical}
    I_E = - \frac{\LR \chi_H}{4G_N} \left( \alpha_0 - \frac{\Delta\phi}{2} \right) = -\frac{c}{6\pi\xi} \left( \alpha_0 - \frac{\Delta\phi}{2} \right)^2.
\end{equation}
Here we have used that the width of the strip is $\Delta x = (2 \alpha_0-\Delta\phi) z_0$. 
 
\subsection{BCFT Spectrum}\label{spectrum}

In this subsection, we examine the spectrum of the BCFT on a strip of width $\Delta x$.
This BCFT is dual to the bulk geometry that is bounded by two EOW branes that are connected at a defect and it also has a conical defect in the bulk. 
For this purpose, we consider the partition function  $Z_{a,b}^{\text{BCFT}}(T_{ \text{BCFT} }^{-1}, \Delta x)$, where $T_{\text{BCFT}}^{-1}$ is the periodicity in $\tau$ direction which we will take to be infinitely large. 
In this limit, we have
\begin{equation}
\label{eq:Z_lim}
    Z_{a,b}^{\text{BCFT}}(T_{\text{BCFT}}^{-1}, \Delta x) \underset{\Delta x\cdot T_{\text{BCFT}} \to 0}{ \xrightarrow{\hspace{2cm}} } e^{-E_{a,b}^{\text{BCFT}}\cdot T_{\text{BCFT}}^{-1}},
\end{equation}
where $E_{a,b}^{\text{BCFT}}$ is the lowest eigenvalue of the BCFT Hamiltonian $H_{a,b}^{\text{BCFT}}$. In general, the corrections to the above equation are exponentially suppressed. Using this lowest eigenvalue, we can define the spectral gap as
\begin{equation}
    \Delta E_{a,b}^{\text{BCFT}} := E_{a,b}^{\text{BCFT}} - \frac{1}{2} E_{a,a}^{\text{BCFT}} - \frac{1}{2} E_{b,b}^{\text{BCFT}} .
\end{equation}

Suppose the EOW branes have dimensionless tensions $\TT_1$ and $\TT_2$. 
Since $Z \approx e^{-I_E}$ and using \eqref{eq:IEconical}, the lowest eigenvalue is
\begin{equation}\label{eq:lowest}
    E_{a,b}^{\text{BCFT}}=-\frac{c}{6\pi \Delta x} \left( \alpha_0 - \frac{\Delta\phi}{2} \right)^2.
\end{equation}
The condition for the boundary changing operator to be the identity operator is
\begin{equation*}
    \alpha_0-\frac{\Delta\phi}{2}=\frac{\pi}{2},
\end{equation*}
so the corresponding lowest eigenvalue is
\begin{equation}
    \label{eq:identity}
    E_{a,b}^{\text{BCFT}}=-\frac{\pi c}{24\Delta x}.
\end{equation}
This situation includes the conventional model where the EOW branes are connected and have identical tensions, and there is no bulk conical defect. 
In particular, when the two boundary conditions are identical, we have $E_{a,a}^{\text{BCFT}} = -\frac{\pi c}{24\Delta x}$.

The lowest eigenvalue of any unitary BCFT must be higher than this value. Therefore, if we demand that the BCFT is unitary, the following condition needs to be satisfied
\begin{equation}
    \label{eq:condition}
    0 \leq \alpha_0 - \frac{\Delta\phi}{2} \leq\frac{\pi}{2}.
\end{equation}
Therefore, for a fixed $\alpha_0$, the spectrum obtained for an arbitrary $\Delta\phi$ subject to \eqref{eq:condition} satisfies
\begin{equation}
    \label{eq:rangealpha0}
    -\frac{\alpha_0^2 c}{6\pi \Delta x} \leq E_{a,b}^{\text{BCFT}} \leq 0.
\end{equation}
We note that \eqref{eq:rangealpha0} generalizes the lowest eigenvalue dictated by the conventional AdS/BCFT model.
In the conventional AdS/BCFT, neither intersections nor interactions between distinct EOW branes are allowed, so such EOW branes are disconnected. 
As the consequence, the lowest eigenvalue is given by
\begin{equation}
\label{eq:disconnected}
    E_{a,b}^{\text{BCFT}} = 0.
\end{equation}
The gravity dual that we have constructed using the defect on EOW branes generalizes this lowest eigenvalue to go below \eqref{eq:disconnected}, for any choice of the two boundary entropies. 
Note that when two boundary conditions are identical, the lowest eigenvalue is given by the identity operator, so the dual geometry corresponds to $\theta_0 = \pi$ with $\Delta\phi = 0$ and $\alpha_0 = \pi/2$.

Finally, we note that $\alpha_0 = \pi/2$ is allowed only if $\TT_1 = \TT_2$, so there is no longer a defect on the EOW branes. 
This implies that we can attain the lowest eigenvalue \eqref{eq:identity} only when the boundary entropy of the two EOW branes are identical.
In this case, the bulk conical defect can reproduce the spectrum
\begin{equation}
    \label{eq:spectrumconical}
    -\frac{\pi c}{24\Delta x}\leq E_{a,b}^{\text{BCFT}}\leq 0.
\end{equation}
For $\TT_1 \neq \TT_2$, the corresponding $\alpha_0$ can be made arbitrarily close to $\pi/2$ by tuning $\theta_0$, but equality cannot be achieved. 

To summarize, a key advantage of our model is that it has a rich spectrum even if two boundary entropies are distinct, as opposed to the conventional AdS/BCFT model which has fixed lowest eigenvalue  \eqref{eq:disconnected}.

\vspace{0.5\baselineskip}
\noindent
\textbf{CFT stress tensor for thermal AdS.}
The above results for the lowest eigenvalue can also be derived by analyzing the metric near the AdS boundary, without knowing the details of the bulk geometry. 
Rewriting the thermal AdS${}_3$ metric in Fefferman-Graham coordinate, we have
\begin{equation}
    \d s_{\text{FG}}^2 = \frac{\LR^2}{r^2}\left(dr^2+ \frac{1+2r^2+r^4}{2z_0^2}\d\tau^2+\frac{1-2r^2+r^4}{2}\d\phi^2\right),
\end{equation}
with $\chi=\frac{2r}{1+r^2}$. Then the dual CFT stress tensor is given by \cite{deHaro:2000vlm}
\begin{equation}
    \langle T_{\tau\tau}\rangle=-\langle T_{xx}\rangle=\frac{\LR}{16\pi G_N}\frac{1}{z_0^2}=\frac{c}{6\pi}\frac{ \left( \alpha_0 - \frac{\Delta\phi}{2} \right)^2 }{(\Delta x)^2}.
\end{equation}
The lowest eigenvalue can be obtained by integrating the CFT stress tensor on the width of the strip,
\begin{equation}
    E_{a,b}^{\text{BCFT}} 
    = - \int_{-\Delta x/2}^{\Delta x/2} \d x \, \langle T_{\tau\tau} \rangle
    = - \frac{c}{6\pi \Delta x} \left( \alpha_0 - \frac{\Delta\phi}{2} \right)^2,
\end{equation}
which matches our previous result.

\vspace{0.5\baselineskip}
\noindent
\textbf{Relation between bulk mass and the spectrum.}
In Euclidean global AdS$_3$, a massive particle at the center is described by the metric
\begin{equation}
    \d s^2 = \left( \frac{r^2}{\LR^2} + 1 - \mu \right)^{-1} \d r^2 + \LR^2 \left( \frac{r^2}{\LR^2} + 1 - \mu \right) \d\tau^2 + r^2 \d\theta^2.
\end{equation}
Here the mass parameter $\mu$ is related to the mass $m$ through
\begin{equation}
    \mu = 8 G_N m
\end{equation}
Also, $\theta$ is periodic with period $2\pi$.

When $\mu < 1$, the massive particle corresponds to a conical defect with deficit angle $\Delta\theta = 2\pi\sqrt{1- \mu} $.
When  $\mu > 1$, this geometry is a Euclidean BTZ black hole.
The relation between $\mu$ and the lowest energy eigenvalue can be obtained from \eqref{eq:lowest}.
In particular, when the EOW brane has no defect i.e., $\alpha_0 = \pi/2$,
\begin{equation}
    2 \sqrt{1 - \mu} - 1 = \sqrt{1 - \frac{24 h_{(a,b)}^{\text{bcc}}}{c}},
\end{equation}
where $h_{(a,b)}^{\text{bcc}}$ is the chiral conformal dimension of the boundary condition changing operator. 
This relation perfectly matches the corresponding relation found in \cite{Kawamoto:2022etl}.

\subsection{Spectral Gap in Liouville Theory}

In this section, we show that Liouville theory with ZZ boundaries \cite{Zamolodchikov:2001ah} gives a spectral gap $\Delta E_{a,b}^{\text{BCFT}}$ which is similar to that of our gravity dual. 
We emphasize that this match is only a formal analogy because ZZ boundaries in Liouville theory have very different properties compared to the usual conformal boundaries of unitary BCFTs.

For comparison, the spectral gap for our gravity model without bulk conical defects is
\begin{equation}
    \Delta E_{a,b}^{\text{BCFT}} = \frac{\pi c}{24\Delta x} \left( 1 - \frac{\alpha_0^2}{(\pi/2)^2} \right) \in \left[ 0, 
    \frac{\pi c}{24\Delta x} \right].
\end{equation}

The central charge for the Liouville theory is $c= 1 + 6 Q^2$ where $Q = b + b^{-1}$. 
The semiclassical limit $c \to \infty$ can be obtained by taking $b \to 0$.
The degenerate representations appear at conformal dimensions 
\begin{equation}
    \Delta_{(m,n)} := \frac{Q^2}{4} - \frac{ ( m/b + nb )^2}{4}
\end{equation}
and the corresponding degenerate characters are
\begin{equation}
    \chi_{m,n}(\tau) = \frac{ q^{-(m/b+nb)^2/4} - q^{-(m/b-nb)^2/4} }{ \eta(\tau) },
    \quad\quad\quad\quad\quad\quad
    q := e^{2\pi i\tau} ,
\end{equation}
where $n,~m$ are positive integers. $\Delta_{(m,n)}$ are negative except for $\Delta_{(1,1)}=0$. Therefore there is no direct connection between Liouville ZZ boundary states and our holographic construction. Nevertheless, when we restrict our attention to the spectral gap $\Delta E_{a,b}^{\text{BCFT}}$, we can find an interesting formal match.

The inner product between two ZZ boundary states $|B_{(m,n)}\rangle$ is given by \cite{Zamolodchikov:2001ah}
\begin{equation}
    \langle B_{(m,n)}| e^{-\beta H/2} |B_{(m',n')} \rangle = \sum_{k=0}^{\min(m,m')-1} \sum_{l=0}^{\min(n,n')-1} \chi_{m+m'-2k-1,n+n'-2l-1}
    \left( \frac{\i}{\beta T_{\text{BCFT}}} \right).
\end{equation}
When we take the limit $\beta T_{\text{BCFT}} \to 0$, the term with $k = l = 0$ dominates, so
\begin{equation}
    \frac{\langle B_{(m,n)}|e^{-\Delta x H}|B_{(m',n')}\rangle}
    {\sqrt{ \langle B_{(m,n)} | e^{-\Delta x H} | B_{(m,n)} \rangle \langle B_{(m',n')} | e^{-\Delta x H} | B_{(m',n')} \rangle} }
    \underset{\Delta x\cdot T_{\text{BCFT}} \to 0}{ \xrightarrow{\hspace{2cm}} }
    e^{-\Delta E_{(m,n), (m',n')}^{\text{BCFT}}
    \cdot T_{BCFT}^{-1} },
\end{equation}
where
\begin{equation}
    \Delta E_{(m,n), (m',n')}^{\text{BCFT}} := \frac{\pi}{4\Delta x} \left( (n-n')b + \frac{m-m'}{b} \right)^2.
\end{equation}

We consider $b \ll 1$, which corresponds to the large $c$ limit.  
Assuming that $m=m'$ and $|n-n'|\leq 1/b^2+1$, the spectral gap ranges between 
\begin{equation}
    0 \leq \Delta E_{(m,n),(m,n')}^{\text{BCFT}} \leq \frac{\pi (c-1)}{24\Delta x} \approx \frac{\pi c}{24\Delta x}.
\end{equation}
This result matches the spectral gap in our AdS/BCFT model. 
However, we should reemphasize that match between Liouville theory with ZZ boundaries and our model is only formal which is clear from $\Delta_{(m,n)}$ being negative.  
Moreover, there is another peculiar feature about ZZ boundaries that distinguishes them from usual BCFT boundaries. 
Namely, we have 
\begin{equation}
    E_{(m,n),(m,n)}^{\text{BCFT}} = \frac{\pi }{4\Delta x}
    \left( (2n-1)b + \frac{2m-1}{b} \right)^2,
\end{equation}
which is distinct from \eqref{eq:identity} although this amplitude is between identical states.
This mismatch also disallows us from interpreting the Liouville theory with ZZ boundaries as the usual BCFT.

\section{Entanglement Entropy and Entanglement Island} \label{section:Entropy}

In this section, we study the holographic entanglement entropy in our model.
We use the Ryu–Takayanagi (RT) prescription to compute this entanglement entropy.
We restrict to the case with no bulk conical defect. 

We start by considering a normalized boundary state for a conformal boundary condition $B_a$ prepared on a circle with circumference $T_{\text{BCFT}}^{-1}$,
\begin{equation}
    | B_a(\Delta x/2) \rangle =
    \frac{e^{-\frac{1}{2} H_{\text{CFT}} \Delta x }|B_a\rangle}
    {\sqrt{\langle B_a|e^{- H_{\text{CFT}} \Delta x }|B_a\rangle}}.
\end{equation}
This state is often used to model an initial state for a global quantum quench\cite{Calabrese:2005in, Calabrese:2009qy}.

In the following, we consider the entanglement entropy of a superposition state
\begin{equation}
    |\Psi \rangle := \sum_a c_a |B_a(\Delta x/2) \rangle,
\end{equation}
where the $c_a$'s are complex coefficients.
We restrict to the case where the boundary entropy $S_{B_a}$ is equal to a fixed value $S_B$ (or it is in a narrow window around $S_B$) for the boundary conditions $B_a$ appearing in $|\Psi \rangle$.
Let $N_{S_B}$ be the number of such boundary states. 

We assume that the inner products $\langle B_a(\Delta x/2)|B_b(\Delta x/2) \rangle$ have holographic duals with a defect connecting the EOW  branes.
For simplicity, we assume that $\theta_{(a,a)} = \pi$ and  $\theta_{(a,b)} = \theta_0$ when $B_a \neq B_b$.
Under these assumptions, the geometries with $\theta_{(a,b)} = \theta_0$ dominate the gravitational computation for $\langle B_a(\Delta x/2)|B_b(\Delta x/2) \rangle$ if
\begin{equation}
    \sum_{a=1}^{N_{S_B}} |c_a|^2 \ll 
    \left( \sum_{ \substack{a,b = 1 \\ a \neq b} }^{N_{S_B}}  c_a^{*} c_b \right) \exp \left[ - \frac{c}{6\pi \Delta x} \left( \frac{\pi^2}{4} - \alpha_0^2 \right) \right].
\end{equation}
If all the $c_a$'s are approximately equal, then this condition can be satisfied only if $N_{S_B} \gg \exp \left[ \frac{c}{6\pi \Delta x} \left( \frac{\pi^2}{4} - \alpha_0^2 \right) \right]$.
Since the gravity dual for $\langle\Psi|\Psi\rangle$ is given by the geometry with $\theta_{(a,b)} = \theta_0$ in this case, the RT surfaces in this geometry compute the entanglement entropy of the state $|\Psi\rangle$.

One can also understand this RT surface in terms of the \emph{pseudo entropy} \cite{Nakata:2020luh}.
The pseudo entropy is given by the von Neumann entropy $S_P(A) = - \Tr [ X_A \log X_A ]$ of a normalized transition matrix $X_A$.
In our case, this transition matrix is
\begin{equation}
    X_A := \frac{ \Tr_A \big[ |B_a(\Delta x/2) \rangle \langle B_b(\Delta x/2)| \big] }{ \Tr_{A A^c} \big[ |B_a(\Delta x/2) \rangle \langle B_b(\Delta x/2)| \big] }.
\end{equation}

\subsection{Three Phases of the RT Surface}

In this subsection, we will use the RT surface prescription to compute the entanglement entropy for a subregion \cite{Calabrese:2005in, Calabrese:2009qy} of the state $|\Psi\rangle$ or the pseudo entropy for $X_A$.
We assume the tensions of the EOW branes are equal, $T_1 = T_2 = T$, and $\TT$ is the corresponding dimensionless tension.
Also, the internal angle at the defect is $\theta_0$. 
We assume the bulk conical defect is absent for simplicity. 

\begin{figure}
    \centering
\begin{subfigure}[b]{0.32\textwidth}
    \centering
    \includegraphics[scale=0.6]{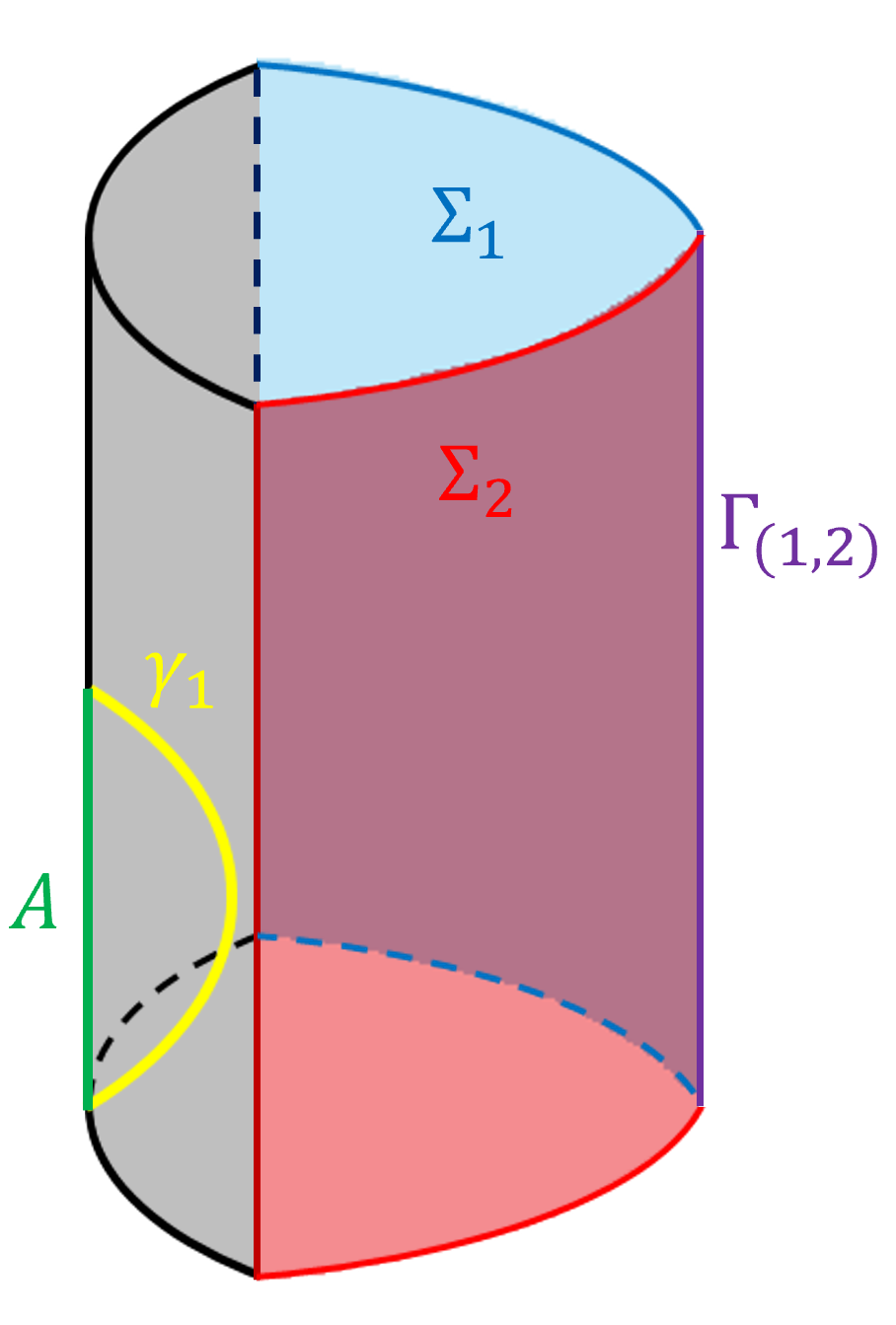}
    \caption{Case 1: Thermal Phase}
    \label{fig:RT1}
\end{subfigure}
\hfill
\begin{subfigure}[b]{0.32\textwidth}
    \centering
    \includegraphics[scale=0.6]{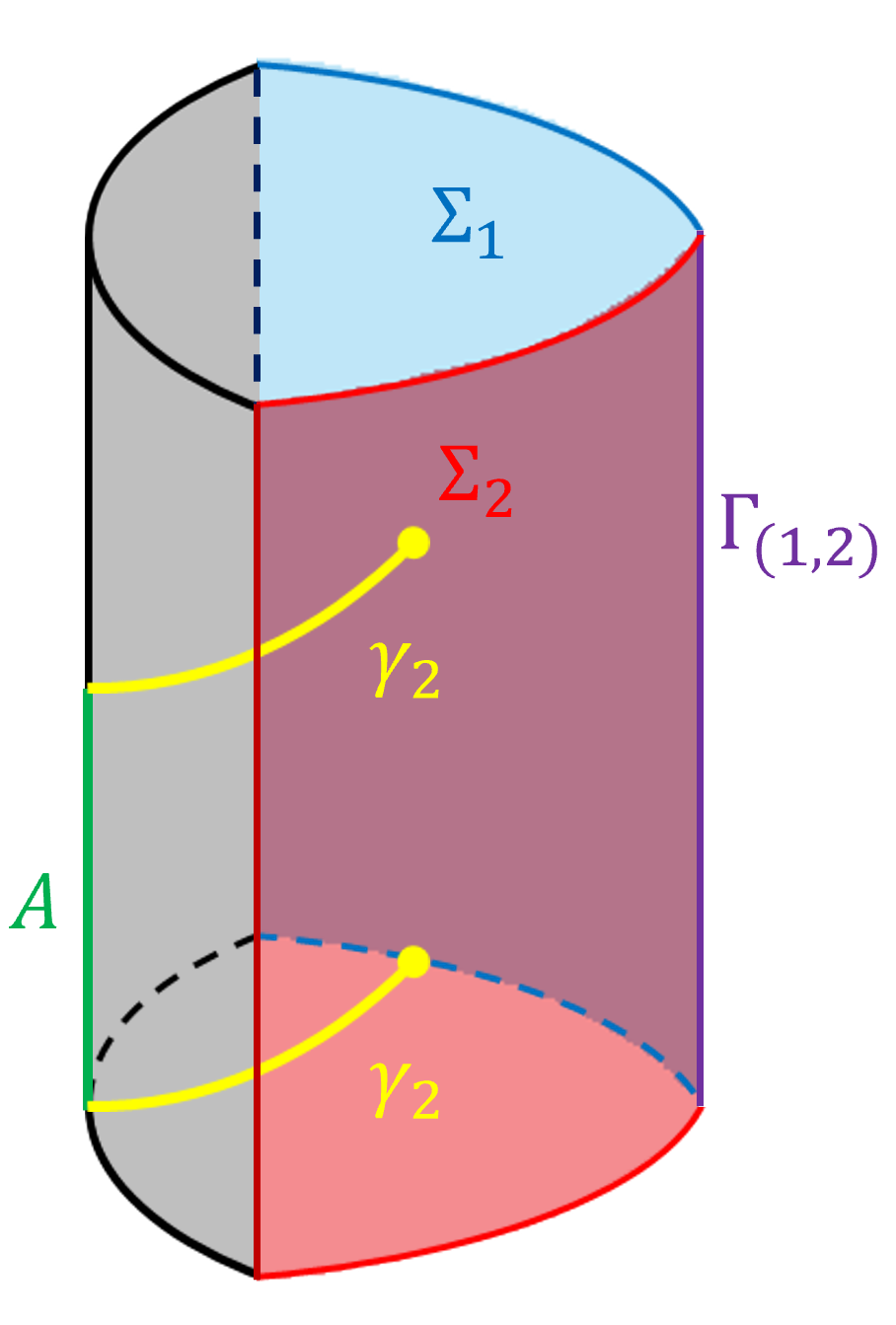}
    \caption{Case 2: Boundary Phase}
    \label{fig:RT2}
\end{subfigure}
\hfill
\begin{subfigure}[b]{0.32\textwidth}
    \centering
    \includegraphics[scale=0.6]{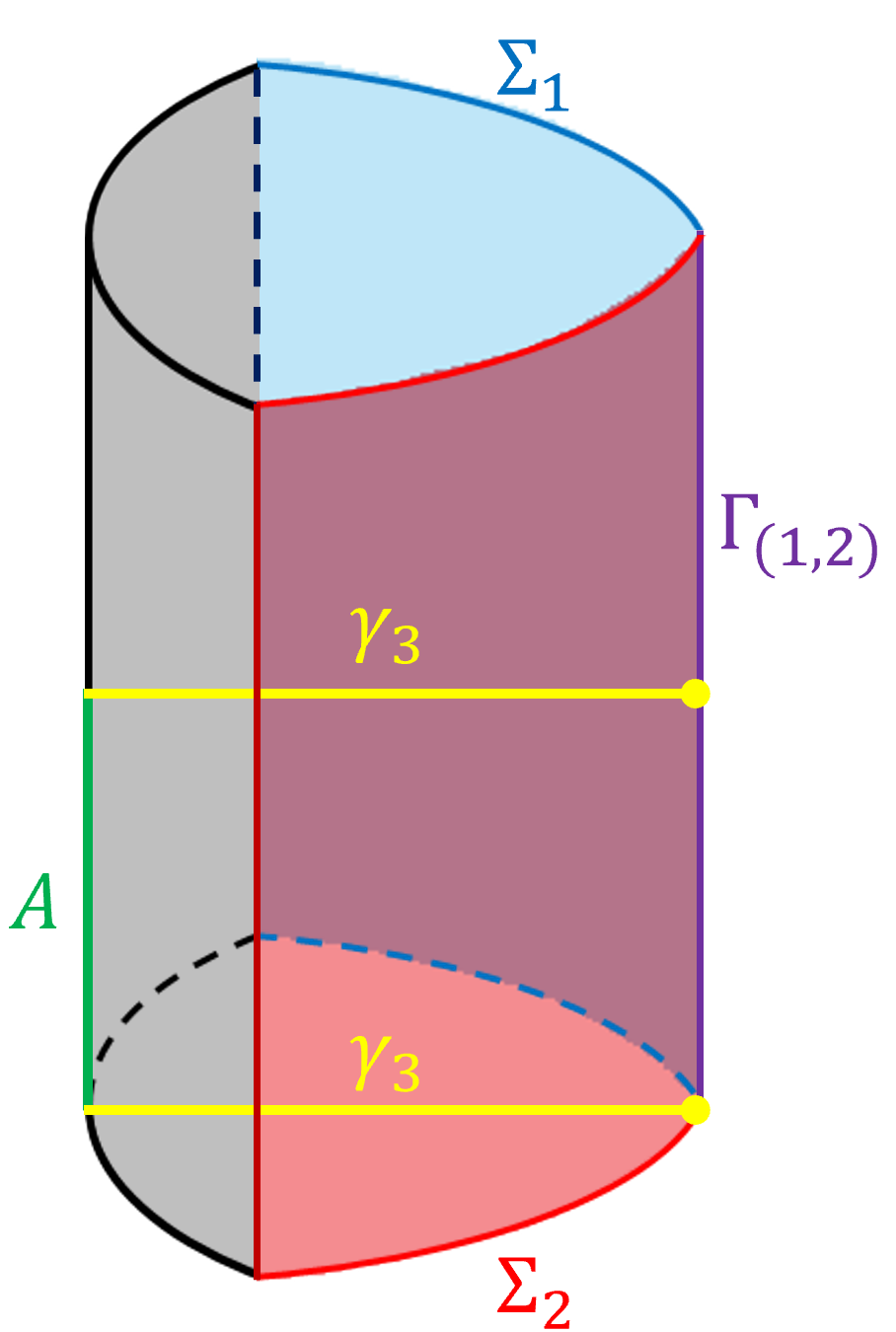}
    \caption{Case 3: Defect Phase}
    \label{fig:RT3}
\end{subfigure}
    
    \caption{Three phases of the RT surfaces for the boundary subregion $A$.}
    \label{fig:RT}
\end{figure}

Recall that our BCFT is defined on the strip 
\begin{equation}
    N = \{ (\tau,\, x = z_0 \phi) ~ | ~ 0 \leq \tau \leq T^{-1}_{\mathrm{BCFT}},~ \pi-\alpha_0 \leq \phi \leq  \pi+\alpha_0\},
\end{equation} 
so the width of the strip is $\Delta x = 2 z_0 \alpha_0$. 
On this strip, we consider the subregion
\begin{equation}
    \rA = \{ (\tau,\, x = z_0(\pi - \sigma) ) ~ | ~ 0 \leq \tau \leq |A| \},
\end{equation}
with a particular value of $\sigma \in [0,\alpha_0)$.
Since $\tau$ is periodic with periodicity $T^{-1}_{\mathrm{BCFT}}$ and $|\Psi\rangle$ is a pure state, it is sufficient to consider the case $|A|<T^{-1}_{\mathrm{BCFT}}/2$.
There are three different phases of the RT surface as shown in Fig.~\ref{fig:RT}.

\vspace{0.5\baselineskip}
\noindent
\textbf{Case 1: Thermal phase.}
In this case, the RT surface is a single connected surface.
This surface is shown in Fig.~\ref{fig:RT1} and we label it as $\gamma_1$.
This phase is realized when the subregion $A$ is sufficiently small.
We  will see that the entanglement entropy is extensive in this case.

The bulk geometry has reflection symmetry about the $\phi = \pi-\sigma$ plane, so the $\gamma_1$ lies on this plane. 
This surface is given by the geodesic
\begin{equation}
    \sqrt{1-\chi^2} = \sqrt{1-\chi_0^2} \cosh \text{\small $ \frac{\tau-\tau_0}{z_0} $}
\end{equation}
with the parameters being
\begin{equation}
    \tau_ 0 = \frac{|A|}{2},
    \quad\quad
    \chi_0 = \tanh \frac{|A|}{2z_0}.
\end{equation} 

Note that $(\tau,\chi,\phi) = (\tau_0,\chi_0,\pi-\sigma)$ is the deepest point on this geodesic.
If the maximum depth $\chi_0$ is sufficiently large, $\gamma_1$ will be cut by the EOW branes in the negative tension case.
Therefore, the thermal phase with $\TT<0$ exists only if $\TT \geq - \frac{ \tan(\alpha_0-\sigma) \sech (|A|/2z_0) }{ \sqrt{ \tan(\alpha_0-\sigma)^2 + \tanh^2 (|A|/2z_0) }}$.

The holographic entanglement entropy is 
\begin{equation}
    S(A) = \frac{\mathcal{A} (\gamma_1)}{4G_N} = \frac{c}{3} \log\frac{2}{\epsilon} +\frac{c}{3} \log\sinh (|A|/2z_0) .
\end{equation}
Since this result is independent of $\sigma$, analytically continuing it to real time does not give rise to any time dependence.
For large $|A|$, this entropy demonstrates the volume law
\begin{equation}
    S(A) \approx \frac{c}{3} \log\frac{2}{\epsilon} + \frac{c|A|}{6z_0} - \frac{c}{3} \log 2,
\end{equation}
at an inverse effective temperature, \begin{equation}
\label{eq:efftemp}
    \beta_{|\Psi \rangle} = 2\pi z_0 = \frac{\pi}{2\alpha_0}(2\Delta x).
\end{equation}
There is a prefactor $\frac{\pi}{2\alpha_0}$ here, which \emph{lowers} the effective temperature for the same $\Delta x$ for a given $\alpha_0 \leq \pi/2$ as compared to the conventional case for which $\alpha_0 = \pi/2$. 

\vspace{0.5\baselineskip}
\noindent
\textbf{Case 2: Boundary phase.}
In this phase, the RT surface has two disconnected pieces that end on the EOW branes.
This surface is shown in Fig.~\ref{fig:RT2} and we label it as $\gamma_2$.
This phase is preferred over the thermal phase for sufficiently large $|A|$.
We will see that the entanglement entropy is intensive in this case.

Each piece of $\gamma_2$ lies on a constant $\tau$ surface because of the $\tau$ reflection symmetry.
These pieces are given by the geodesic
\begin{equation}
    \sqrt{1-\chi^2} \cos(\phi - \pi + \sigma + \phi_0) = \cos\phi_0.
\end{equation}
Here we have determined the parameter $\phi_0 = \alpha_0 - \sigma$ by requiring that the geodesic intersects the EOW brane orthogonally.
The endpoints of $\gamma_2$, lying on the EOW brane $\Sigma_1$, are given by $\chi_{\text{ep}} = \sqrt{1-\TT^2} \sin(\alpha_0-\sigma)$.

If $\alpha_0 \leq \pi/2$, or equivalently, $\theta_0 \leq \pi$, this geodesic exists for all values of $\sigma$. 
If $\alpha_0 \geq \pi/2$, the critical value corresponds to the endpoint lying on the defect.
This is given by $\chi_{\text{ep}} = \chi_* = \sqrt{1-\TT^2\csc^2(\frac{\theta_0}{2})}$, so the condition for the existence of this geodesic is
\begin{equation}
\label{eq:conditioncase2largealpha}
    \alpha_0 - \sigma \leq \acos \text{\small $ \frac{ \TT \left|\cot (\frac{\theta_0}{2}) \right| }{ \sqrt{1-\TT^2} } $}.
\end{equation}
In particular, this surface exists for $\sigma=0$ if and only if $\alpha_0 \leq \pi/2$.

The holographic entanglement entropy is
\begin{equation}
    S_A=\frac{\mathcal{A}(\gamma_2)}{4G_N} = \frac{c}{3} \log\frac{2}{\epsilon}+ \frac{c}{3} \log \text{\small $ \sin(\alpha_0 - \sigma) $} + 2S_B.
\end{equation}
Here $S_B$ is the boundary entropy
\begin{equation}
    S_B=\frac{c}{6} \atanh \TT,
\end{equation}
which can also be obtained using the disk partition function \cite{Fujita:2011fp}.
Note that there is an extremal but not minimal surface $\gamma_2'$ ending on the other EOW brane $\Sigma_2$, i.e., the EOW brane anchored at $\phi = \pi + \alpha_0$. 
For this surface, we have
\begin{equation}
    \frac{\mathcal{A}(\gamma_2')}{4G_N} = \frac{c}{3} \log\frac{2}{\epsilon}+ \frac{c}{3} \log \text{\small $ \sin(\alpha_0 + \sigma)$} + 2S_B.
\end{equation}
The surface $\gamma_2'$ exists when $\alpha_0 + \sigma \leq \acos \text{\small $ \frac{ \TT \left|\cot (\frac{\theta_0}{2}) \right| }{ \sqrt{1-\TT^2} } $} $. 

When we analytically continue this Euclidean entanglement entropy to real time, it is important to consider both $\mathcal{A}(\gamma_2)$ and $\mathcal{A}(\gamma_2')$, as well as cases where the two pieces of the RT surface ends on distinct EOW branes. To evaluate the entanglement entropy of the state $e^{-\i H^{\text{CFT}} t} |\Psi\rangle$, we analytically continue  $\frac{\mathcal{A}(\gamma_2)}{4G_N}$ and $\frac{\mathcal{A}(\gamma_2')}{4G_N}$ by taking $\sigma \to -\i \frac{t}{z_0}$.
After this substitution, these two extremal areas are complex conjugates of the other, $\mathcal{A}(\gamma_2) = \overline{\mathcal{A}(\gamma_2')}$.
The total areas in the other two cases are purely real and are equal to $\Re{ \mathcal{A} (\gamma_2)} = \Re{ \mathcal{A} (\gamma_2')}$.
Therefore, the real time entanglement entropy is given by this real piece,
\begin{equation}
    S_A = \Re{ \frac{\mathcal{A} (\gamma_2)}{4G_N}} =\frac{c}{3} \log\frac{2}{\epsilon}+ \frac{c}{6} \log \text{\small $ \frac{\cosh(2t/z_0) - \cos(2\alpha_0) }{2} $} + 2S_B - \frac{c}{3} \log 2.
\end{equation}
At late times $t \gg z_0$, this entanglement entropy is
\begin{equation}
    S_A \approx \frac{c}{3} \log\frac{2}{\epsilon} + \frac{ct}{3z_0} + 2S_B.
\end{equation}
Hence, at late times satisfying 
\begin{equation}
   t>\frac{|A|}{2}-\frac{6z_0S_B}{c},
\end{equation}
the thermal phase will be favored over the boundary phase.

\vspace{0.5\baselineskip}
\noindent
\textbf{Case 3: Defect phase.}
In this phase, the RT surface has two disconnected pieces that end on the defect.
This surface is shown in Fig.~\ref{fig:RT3} and we label it as $\gamma_3$.
As in the boundary phase, the entropy is intensive. 
This phase is always subdominant compared to the boundary entropy phase if the latter exists.
Therefore, this phase is realized only when the subregion is sufficiently large and the boundary entropy phase is absent. 
For this reason, when $\alpha_0 \leq \pi/2$, this phase is irrelevant.

As earlier, each piece of $\gamma_3$ lies on a constant $\tau$ surface.
These pieces are given by the geodesic
\begin{equation}
    \sqrt{1-\chi^2} \cos(\phi - \pi + \sigma + \phi_0) = \cos\phi_0,
\end{equation}
Here we have determined the parameter $\phi_0  = \atan \frac{\sin(\theta_0/2)  + \TT \cos\sigma}{\TT \sin\sigma}$ by requiring that the geodesic passes through the defect.

If $\TT>0$ this geodesic is guaranteed to exist.
However, when $\TT<0$ this geodesic exists only if $\sigma \leq 2 \atan \left( \text{\small $ \tan \left( \frac{\theta_0}{4} \right)  \sqrt{\frac{ \sin(\theta_0/2)+\TT }{ \sin(\theta_0/2)-\TT }} $}  \right)$.
This implies that this phase always exists around $\sigma=0$. 

The holographic entanglement entropy is
\begin{equation}
    S_A = \frac{\mathcal{A}(\gamma_3)}{4G_N} = \frac{c}{3} \log\frac{2}{\epsilon}+ \frac{c}{3}\log \text{\small $ \frac{\sin(\theta_0/2) + \TT \cos\sigma}{\sin(\theta_0/2) + \TT} $} + 2S_D.
    \label{eq:defectentropy}
\end{equation}
Here we have defined
\begin{equation}
    S_D :=  \frac{c}{12} \log \text{\small $ \frac{ \sin(\theta_0/2) + \TT }{ \sin(\theta_0/2) - \TT } $},
\end{equation}
and we call it the defect entropy. 
This defect entropy can be considered as a generalization of the boundary entropy $S_B$ and it reduces to $S_B$ if we make the defect disappear by setting $\theta_0 = \pi$.

As mentioned earlier, the boundary phase is favored over this phase, $\mathcal{A}(\gamma_2) < \mathcal{A}(\gamma_3)$, if both phases exist.
This is because $\gamma_2$ is minimal amongst all surfaces that end on the EOW brane. 

The real time entanglement entropy is given by analytically continuing $\sigma \to -\i \frac{t}{z_0}$, so
\begin{equation}
    S_A = \frac{c}{3} \log\frac{2}{\epsilon}+ \frac{c}{3}\log \left(\text{\small $ \frac{ \TT \cosh(t/z_0) + \sin(\theta_0/2)  }{\TT + \sin(\theta_0/2) } $}\right) + 2S_D .
\end{equation}
Note that this result is manifestly real unlike case 2. 
At late times $t \gg z_0$, this entanglement entropy is
\begin{equation}
    S_A \approx \frac{c}{3} \log\frac{2}{\epsilon} + \frac{ct}{3z_0} + 2S_D - \frac{c}{3} \log \text{\small $ \frac{2 \left( \TT + \sin(\theta_0/2) \right) }{\TT} $} .
\end{equation}

It is also worth mentioning that the $\gamma_3$ is minimal but not extremal because of the singular nature of the defect.
One can imagine a scenario when there is a matter theory on the EOW brane which smoothly interpolates between the two different tensions of the EOW branes, and the defect is realized as the sharp limit.
In this case, the RT surface is extremal and the distinction between cases $2$ and $3$ disappears.

\subsection{Interpretation as Entanglement Island}

In this subsection, we explain and interpret the results in the previous section in terms of the island formula.
We interpret the state $|\Psi\rangle$ as a field theoretic wavefunction being prepared by a matter field that lives on the EOW brane.
In other words, the EOW brane is now the spacetime on which a quantum state $|\Psi\rangle$ is prepared \cite{Chen:2020tes, Miyaji:2021lcq}.
We are interested in the entanglement entropy of a subregion $A$ of the state $|\Psi\rangle$.
We assume that the matter theory is a holographic BCFT to simplify the analysis \cite{Sully:2020pza}.
When the effective temperature of $|\Psi\rangle$ is sufficiently low, the EOW branes in the bulk are disconnected.
When the temperature is sufficiently high, the EOW branes are connected via a defect, creating a closed universe that terminates at this defect. 

The three phases of the RT surface in the previous section can be interpreted as the phases of the entanglement entropy of $A$ \cite{Miyaji:2021lcq}. 
When the subregion is sufficiently small, the entanglement entropy is given by the thermal answer from matter theory.
This corresponds to case $1$. 
When the subregion is large enough so that the naive thermal entropy is much larger than the boundary entropy, then an entanglement island is formed on the EOW brane (case $2$) or on the defect (case $3$).

In case $2$, the two pieces of the RT surface can end on the same EOW brane or on different EOW branes.
However, the ``island'' in the latter case cuts through the defect on the EOW brane, so the corresponding entropy increases by the boundary entropy of the defect coming from the matter field on the EOW branes. 
Consequently, it can be argued that this configuration is subdominant when we are looking for the island.
As the result, the island lies only on one of the two EOW branes. 
The bulk matter wavefunction on such an island is a transition matrix, so the matter part of the generalized entropy is given by the pseudo entropy.
It is expected that there is a generalization of entanglement wedge reconstruction to such cases, although there are no quantum information theoretic foundations for this claim yet.

In case $3$, the RT surface ends on the defect.
This can be interpreted as an island lying exclusively on the defect.
Therefore, a portion of the defect is included in the entanglement wedge of the subregion $A$.

\section{Three-dimensional Wormhole Saddles}
\label{section:Replica}

In this section, we construct a connected bulk geometry with multiple AdS boundaries using our model of EOW branes and defects. 
This geometry is similar to the replica wormhole in $2$d gravity with EOW branes \cite{Penington:2019kki}.
However, we will be interested in defects with $\theta_0 > \pi$, so these defects have negative energy.

One way to add such negative energy in the bulk is by adding non-local interaction between different boundaries \cite{Gao:2016bin}, which can originate from integrating out ``fast'' degrees of freedom.
Such non-local interaction between boundaries can make a non-traversable Einstein-Rosen bridge traversable, so that different boundaries can communicate with each other \cite{Gao:2016bin, Maldacena:2018lmt}.
We interpret our model in this section as a realization of such non-local interaction between BCFTs, and the non-unitarity of the BCFTs with $\theta_0>\pi$ comes from such non-local interactions obtained by integrating out these fast degrees of freedom.

\begin{figure}
    \centering
\begin{subfigure}[b]{0.32\textwidth}
    \centering
    \includegraphics[scale=0.6]{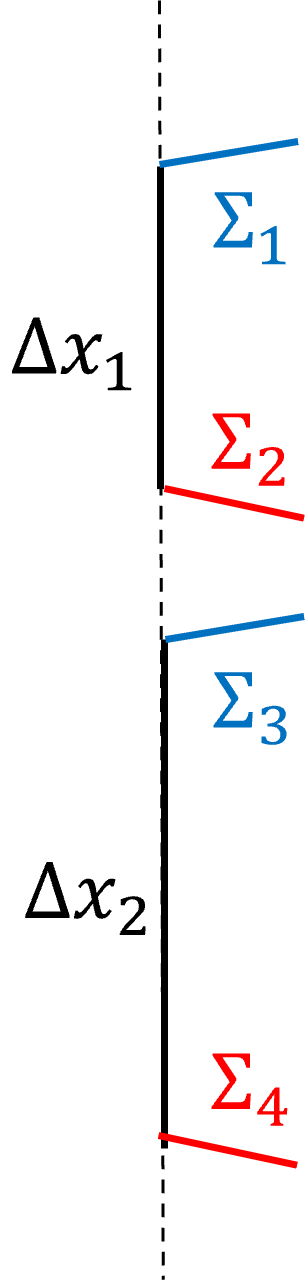}
    \caption{Boundary Setup}
    \label{fig:FW1}
\end{subfigure}
\hfill
\begin{subfigure}[b]{0.32\textwidth}
    \centering
    \includegraphics[scale=0.6]{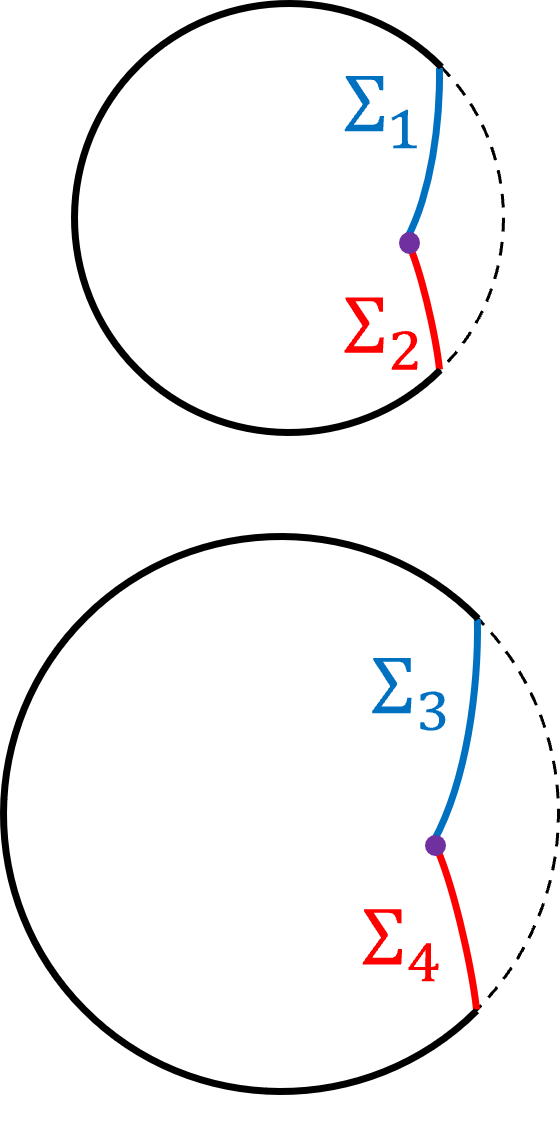}
    \caption{Factorized Geometry}
    \label{fig:FW2}
\end{subfigure}
\hfill
\begin{subfigure}[b]{0.32\textwidth}
    \centering
    \includegraphics[scale=0.6]{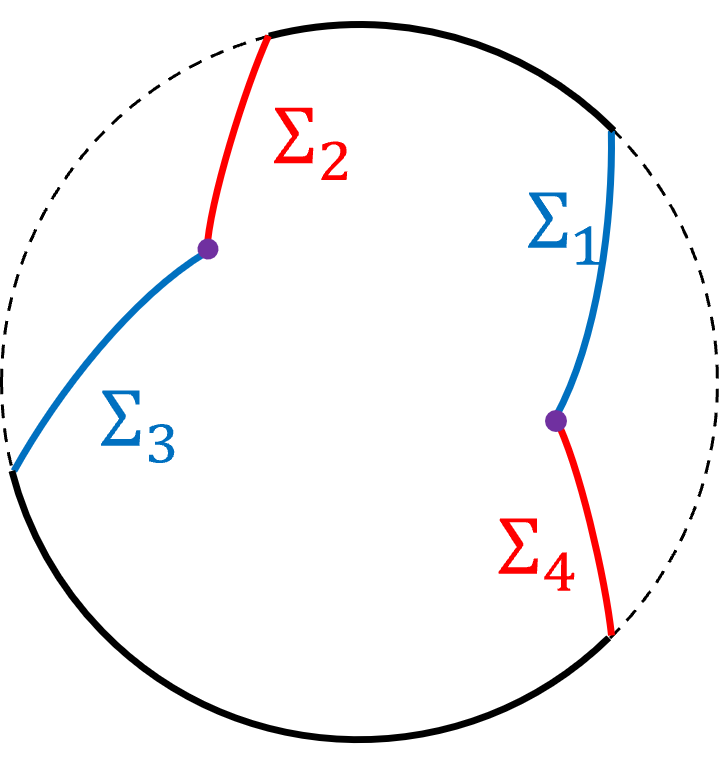}
    \caption{Wormhole Geometry}
    \label{fig:FW3}
\end{subfigure}
    
    \caption{Constant $\tau$ slices showing two BCFTs along with the disconnected and connected dual geometries. The factorized geometry is the leading contribution to the product of partition functions whereas the wormhole geometry is subleading. The latter arises from non-unitarity of the BCFTs with $\theta_0 > \pi$.}
    \label{fig:FW}
\end{figure}

Let us consider two BCFTs that live on two different strips with widths $\Delta x_1$ and $\Delta x_2$ respectively.
They have the same periodicity $T_{\mathrm{BCFT}}^{-1}$ in $\tau$ direction, so the aspect ratios are $\xi_1 = \Delta x_1 \cdot T_{\mathrm{BCFT}}$ and $\xi_2 = \Delta x_2 \cdot T_{\mathrm{BCFT}}$.
Fig.~\ref{fig:FW1} shows a constant $\tau$ slice for this setup.

We assume that the tensions satisfy $T_1 = T_3$ and $T_2 = T_4$ and that the angle of intersection between the $\Sigma_1, \, \Sigma_3$ EOW branes and the $\Sigma_2, \, \Sigma_4$ EOW branes is $\theta_0 > \pi$. 
This assumption is required to get connected geometries.
it implies that $\alpha_0 > \frac{\pi}{2}$ and that the BCFTs are non-unitary.

The factorized geometry is shown in Fig.~\ref{fig:FW2}.
The Euclidean action for this geometry is the sum of the two corresponding actions
\begin{equation}
    I_E^{\mathrm{f}} =  - \frac{c}{6\pi}  \frac{ \alpha_0^2 }{\xi_1} - \frac{c}{6\pi} \frac{ \alpha_0^2 }{\xi_2}.
\end{equation}
where $\alpha_0$ is given by \eqref{eq:alpha0}. 
The connected wormhole geometry is shown in Fig.~\ref{fig:FW3}.
It is obtained by stitching together two geometries, each with a boundary angle $2\alpha_0$.
The total boundary angle for this wormhole geometry is $2\alpha_{\mathrm{w}} = (4\alpha_0-2\pi)$, where we have subtracted a $2\pi$ to account for the connection.
The Euclidean action for the wormhole geometry is
\begin{equation}
    I_E^{\mathrm{w}} 
    = - \frac{c}{6\pi \xi_\mathrm{w}} \alpha_\mathrm{w}^2 
    = - \frac{c}{6 \pi} \frac{ (2\alpha_0-\pi)^2 }{\xi_1+\xi_2},
\end{equation}
where $\xi_\mathrm{w} = \xi_1+\xi_2$ is the total aspect ratio for the wormhole.

The factorized geometry strictly dominates over the wormhole geometry, since
\begin{equation}
    I_E^{\mathrm{f}} = - \frac{c \alpha_0^2 }{6 \pi}\left(  \frac{1}{\xi_1} + \frac{1}{\xi_2} \right)
    \leq - \frac{c \alpha_0^2 }{6 \pi} \frac{4}{\xi_1+\xi_2} <
    - \frac{c(2\alpha_0^2 -\pi)^2}{6 \pi} \frac{1}{\xi_1+\xi_2} = I_E^{\mathrm{w}}.
\end{equation}
Here we have used the arithmetic mean-harmonic mean inequality to establish the first inequality. 

More generally, if we have $n$ disconnected boundary regions with aspect ratios $\xi_i = \Delta x_i \cdot T_{\mathrm{BCFT}}$, then the Euclidean actions for the factorized geometry is
\begin{equation}
    I_E^{\mathrm{f}} = - \frac{c}{6 \pi} \sum_{i=1}^{n}  \frac{ \alpha_0^2 }{\xi_i},
\end{equation}
and for the fully connected wormhole geometry is
\begin{equation}
    I_E^{\mathrm{w}}
    = - \frac{c}{6 \pi} \frac{ \left( n\alpha_0 -(n-1)\pi \right)^2 }{ \sum_{i=1}^{n} \xi_i },
    \quad\quad .
\end{equation}
This fully connected geometry exists only for $\alpha_0 \geq \frac{(n-1)\pi}{n}$.
From these results, we can conclude that any wormhole geometry is subdominant compared to the corresponding factorized geometry, even though we have considered a non-unitary BCFT.
Recall that without non-unitarity, it is not even possible to construct a connected saddle. 
Interestingly, in order to increase the number of boundary components, we need larger a $\alpha_0$, which means that we need to make the BCFT ``more'' non-unitary.

\section{Discussion}\label{section:discussion}

In this paper, we have developed a generalization of the conventional AdS/BCFT model by including defects that connect EOW branes with different tensions.
This construction enables us to study a BCFT whose lowest eigenvalue can be tuned arbitrarily close to that of the identity operator.
This construction is particularly useful when the boundary entropies of the two boundaries are distinct, i.e. when the conventional model has a restricted lowest eigenvalue. 

We conclude with some remarks and possible future directions.
The construction in our model is based on $3$d gravity, and we expect that the generalization to higher dimensions should be straightforward.
Although we have calculated the real time entropy in section  \ref{section:Entropy}, we did not study the corresponding real time geometry. 
It would be interesting to develop this analytic continuation.
It would also be interesting to generalize our defect on the EOW branes to a smooth configuration.
We can imagine a model in which the tension of the EOW brane is position dependent, and in particular, it interpolates smoothly between the two BCFT boundaries. Closely related models were studied in \cite{Erdmenger:2014xya, Erdmenger:2015spo}. We would like to understand the interpretation of the boundary entropy in this configuration, for which a specific example is given by the defect entropy in \eqref{eq:defectentropy}.

The connection between unitarity and connected geometries in section \ref{section:Replica} remains mysterious and interesting. 
We should emphasize that although the connected saddle exists in the non-unitary case, this saddle is always subleading compared to the disconnected saddle.

\subsection*{Acknowledgements}
We thank Yuya Kusuki and Tadashi Takayanagi for useful discussions and comments.
C.M. is supported in part by the U.S. Department of Energy, Office of Science, Office of High Energy Physics under QuantISED Award DE-SC0019380 and contract DE-AC02-05CH11231. 

\appendix

\section{Extention of Boundary Conformal Transformation to Bulk}
\label{appendix:conformal_transformation}

In this appendix, we examine the bulk coordinate transformation from the Poincare patch to another coordinate patch.
This transformation is dual to the boundary conformal transformation from a plane to a $2$d patch.
In particular, we find the map between vacuum AdS and thermal AdS.

The Poincare AdS${}_3$ metric is
\begin{equation}
    \d s^2 = \frac{\LR^2}{z^2} \left( -2\d v_{+} \d v_{-} + \d z^2 \right).
\end{equation}
Here we have used the light cone coordinate $v_{\pm} := \frac{1}{\sqrt{2}}(t\pm r)$ for the Minkowski space $\d s^2 = -2\d v_{+} \d v_{-} = -\d t^2 + \d r^2$ at the boundary of AdS.
 
Suppose we have a conformal transformation $v_{\pm} = f_{\pm}(u_{\pm})$ that acts on the boundary.
We can extend this to a bulk coordinate transformation via \cite{Roberts:2012aq} \begin{equation}
\label{eq:bulktransformation}
    \begin{cases}
    v_{\pm} = f_{\pm}(u_{\pm}) +
    \frac{2\zeta^2 f'_{\pm}(u_{\pm})^2 f''_{\mp}(u_{\mp})}
    {8f_{+}'(u_{+}) f_{-}'(u_{-}) - \zeta^2 f_{+}''(u_{+}) f_{-}''(u_{-})} ,\\
    z = \frac{ 8 \zeta (f'_{+}(u_{+}) f'_{-}(u_{-}) )^{3/2}}
    {8f_{+}'(u_{+})f_{-}'(u_{-}) - \zeta^2 f_{+}''(u_{+}) f_{-}''(u_{-})}.
    \end{cases}
\end{equation}
where $\zeta$ is the new radial coordinate.
This bulk transformation maps the Poincare AdS$_3$ metric to
\begin{equation}
    \d s^2 = \LR^2 \left( \frac{\d \zeta^2}{\zeta^2} + L_{+} \d u_{+}^2 + L_{-} \d u_{-}^2 - \left( \frac{2}{\zeta^2} + \frac{\zeta^2}{2} L_{+}L_{-} \right) \d u_{+} \d u_{-} \right).
\end{equation}
Here $L_{\pm}$ are related to the Schwarzian of the conformal transformation and are given by
\begin{equation}
    L_{\pm} := -\frac{1}{2} \{ f_{\pm}(u_{\pm}) , u_{\pm} \} = \frac{3f_{\pm}''^2 - 2f_{\pm}' f_{\pm}'''}{4f_{\pm}'^2}.
\end{equation}
Hence, the CFT stress tensor in the transformed coordinates is
\begin{equation}
    T_{u_\pm u_\pm} =\frac{c}{12\pi} L_{\pm},
    \quad\quad\quad\quad
    T_{u_+ u_-}=0.
\end{equation}
The analytical continuation of $v_{\pm}$ to imaginary time is
\begin{equation}
    v_{\pm} = \frac{t \pm r}{\sqrt{2}} \to \frac{-\i\tau \pm r}{\sqrt{2}}.
\end{equation}
Similarly, for $\d s^2 = -2\d x_+ \d x_- = -\d t_x^2 + z_0^2 \d\phi^2$, we have
\begin{equation}
    u_{\pm} = \frac{t_u \pm z_0\phi}{\sqrt{2}} \to \frac{-\i\tau_u \pm z_0\phi}{\sqrt{2}}.
\end{equation}

In these complexified coordinates, the conformal transformation can be rewritten as $v = r+\i\tau = g(u), ~\bar{v} = r-\i\tau = \bar{g}(\bar{u})$ with $u := z_0 \phi+\i\tau_u, ~\bar{u} := z_0 \phi-\i\tau_u$.
The metric is
\begin{equation}
    \d s^2 = \LR^2 \left( \frac{\d \zeta^2}{\zeta^2} + L \d u^2 + \bar{L} \d\bar{u}^2 + \left( \frac{1}{\zeta^2} + \zeta^2 L\bar{L} \right) \d u \d\bar{u} \right).
\end{equation}
where we have defined $L := \frac{1}{2} \{g(u),u\} = - \frac{1}{4} \{f_-(u_-), u_-\}$ and $\bar{L} := - \frac{1}{2} \{\bar{g}(\bar{u}), \bar{u}\} = - \frac{1}{4} \{f_+(u_+), u_+\}$.

\subsection*{Mapping a Plane to a Cylinder}

We consider a conformal transformation from a cylinder to a plane,
\begin{equation}
    \label{eq:transformation}
    v = z_0 e^{ -\frac{\i}{z_0} u },
    \quad\quad
    \bar{v} = z_0 e^{ \frac{\i}{z_0} \bar{u} }.
\end{equation}
Here we have defined the coordinates on the cylinder as
\begin{equation}
    u := z_0 \phi + \i\tau = -\sqrt{2} u_{-},
    \quad
    \bar{u} := z_0 \phi - \i\tau = \sqrt{2} u_+, 
\end{equation}
and the coordinates on the plane as
\begin{equation}
    v := x - \i y = -\sqrt{2} v_{-},
    \quad
    \bar{v} := x + \i y = \sqrt{2} v_+.
\end{equation}
This conformal transformation can be also be expressed as $v_{\pm} = \pm\frac{z_0}{\sqrt{2}} e^{\frac{\i\sqrt{2}}{z_0} u_{\pm}}$.

Applying \eqref{eq:bulktransformation} to this conformal transformation, we obtain the bulk transformation  between the Poincare metric, $\d s^2 = \frac{\LR^2}{s^2} (\d x^2 + \d y^2 + \d z^2)$, and thermal AdS, $\d s^2 = \LR^2 \left( \frac{\d\tau^2}{z_0^2\chi^2} + \frac{\d\chi^2}{h(\chi)\chi^2} + \frac{h(\chi)\d\phi^2}{\chi^2} \right)$, to be
\begin{equation}
    \begin{cases}
    v_{\pm} = \pm\frac{z_0}{\sqrt{2}} \sqrt{h(\chi)} e^{\i\sqrt{2}u_{\pm}/z_0}, \\
    z = z_0 e^{i(u_+ + u_-)/(\sqrt{2} z_0)} \chi.
    \end{cases}
\end{equation}
where we have related $\chi = \frac{4z_0 \zeta}{\zeta^2 + 4z_0^2}$.
This can also be expressed as
\begin{equation}
\label{eq:coordinatetransformation}
    \begin{cases}
    x=z_0 e^{\tau/z_0} \sqrt{h(\chi)} \cos\phi ,\\
    y=z_0 e^{\tau/z_0} \sqrt{h(\chi)} \sin\phi,\\
    z=z_0 e^{\tau/z_0} \chi,
    \end{cases}
\end{equation}

For this transformation,
\begin{equation}
    L_{\pm}=-\frac{1}{2z_0^2},
\end{equation}
so the stress tensor is given by
\begin{equation}
    T_{\tau\tau}=-\frac{T_{\phi\phi}}{z_0^2}=\frac{c}{24\pi z_0^2}.
\end{equation}

\section*{}

\bibliographystyle{apsrev4-1long}
\bibliography{main}

\end{document}